\documentclass[aps,prb,twocolumn,groupedaddress,amsmath,amssymb,amsfonts]{revtex4}

\usepackage{latexsym,amssymb}
\usepackage{graphicx,color,pstricks}
\usepackage{epsfig,epsf,bm}

\definecolor{gray}{rgb}{0.7,0.7,0.7}

\begin{document}

\title{The phase diagram of a two-dimensional dirty tilted Dirac semimetal}

\author{Yu-Li Lee}
\email{yllee@cc.ncue.edu.tw}
\affiliation{Department of Physics, National Changhua University of Education, Changhua, Taiwan, R.O.C.}

\author{Yu-Wen Lee}
\email{ywlee@thu.edu.tw}
\affiliation{Department of Applied Physics, Tunghai University, Taichung, Taiwan, R.O.C.}

\date{\today}

\begin{abstract}
 We investigate the effects of quenched disorder on a non-interacting tilted Dirac semimetal in two
 dimensions. Depending on the magnitude of the tilting parameter, the system can have either Fermi points
 (type-I) or Fermi lines (type-II). In general, there are three different types of disorders for Dirac
 fermions in two dimensions, namely, the random scalar potential, the random vector potentials along and
 perpendicular to the tilting direction, and the random mass. We study the effects of weak disorder in
 terms of the renormalization group, which is performed by integrating out the modes with large energies,
 instead of large momenta. Since the parametrization of the low-energy degrees of freedom depends on the
 structure of the Fermi surface, the resulting one-loop renormalization-group equations depend on the
 type of tilted Dirac fermions. Whenever the disorder is a marginal perturbation, we examine its role on
 low-energy physics by a mean-field approximation of the replica field theory or the first-order Born
 approximation. Based on our analysis, we suggest the phase diagrams of a two-dimensional tilted Dirac
 fermion in the presence of different types of disorder.
\end{abstract}

\maketitle

\section{Introduction}

The nodal semimetals, including the Dirac and the Weyl fermions, in solid-state materials have attracted
intense theoretical and experimental interests in recent years\cite{Vafek,Wehling}. On account of the
linear quasiparticle dispersion, which results in a vanishing density of states (DOS) at the Fermi level,
and the non-trivial topological properties, these materials exhibit many interesting phenomena that are
different from the ordinary metals described by the Fermi-liquid (FL) theory. Examples of two-dimensional
($2$D) Dirac semimetals (DSMs) include the graphene\cite{CastroNeto,Peres,Kotov} and the surface states
of a three-dimensional ($3$D) topological insulator\cite{Hasan,XLQi}. Very recently, the Weyl semimetals
(WSMs) have also been detected experimentally in the non-centrosymmetric but time-reversal preserving
materials, such as TaAs, NbAs, TaP, and NbP\cite{CShekhar,BQLv,SYXu,BLv,LYang,SYXu2,NXu}.

Due to the lack of a fundamental Lorentz symmetry, the spectra of DSMs (WSMs) realized in solid-state
materials do not have to be isotropic. In particular, they can be tilted\cite{AASoluyanov}. Depending on
the magnitude of the tilting angle, there are two types of DSMs (WSMs). For type-I DSMs (WSMs), the Dirac
(Weyl) cone is only moderately tilted such that they still have a point-like Fermi surface at the Dirac
(Weyl) node. When the tilting angle is large enough, the electron and hole Fermi surfaces can coexist
with the band-touching Dirac/Weyl nodes. This leads to a new kind of materials, which are commonly
referred to as type-II DSMs (WSMs)\cite{AASoluyanov}. In three dimensions, the tilted Weyl cones were
proposed to be realized in a material WTe$_2$\cite{AASoluyanov}, a spin-orbit coupled fermionic
superfluid with the Fulde-Ferrell ground state\cite{Xu1}, or a cold-atom optical lattice\cite{Xu2}. On
the other hand, in two dimensions, the tilted Dirac cones were proposed to be realized in a mechanically
deformed graphene and the organic compound $\alpha$-(BEDT-TTF)$_2$I$_3$\cite{Katayama,Kobayashi,MOGoerbig}.
Recently, type-II Dirac fermions are experimentally discovered in two materials: PdTe$_2$\cite{HJNoh,FCFei}
and PtTe$_2$\cite{MZYan}.

Since the disorder is ubiquitous in condensed-matter systems, its role on the nodal semimetals is an
interesting topic from both the theoretical and experimental perspectives. For the non-tilting case, in
a pioneering work\cite{Fradkin}, Fradkin showed that unlike the usual FL, a $3$D DSM is stable against
the presence of a weak random scalar potential. When the disorder strength is beyond some critical value,
there is a quantum phase transition (QPT) which separates the DSM from a diffusive metal (DM) with a
non-zero DOS at the Fermi level. In two dimensions, the system behaves more like an ordinary disordered
FL. That is, the ground state is always localized so that the system is an insulator.

The effects of a random scalar potential on the type-I WSM have been studied in Refs. \onlinecite{Trescher}
and \onlinecite{Sikkenk}. The results are similar to those of the untilted case. That is, the semimetallic
phase remains stable for weak disorder. However, the presence of tilt decreases the region occupied by
the semimetallic phase due to the reduction of the critical disorder strength for the QPT to the DM. In
the mean time, the disorder increases effective tilt of the quasiparticle excitations in the semimetallic
phase.

In the present paper, we would like to study the ground state of a non-interacting $2$D tilted Dirac
fermions in the presence of quenched disorder. We adopt the renormalization-group (RG) method which is
performed by integrating out disorder at each order in the perturbation theory. It is known from the
study of the FL theory that the RG transformation must scale toward the Fermi surface, instead of the
origin in the momentum space\cite{Shankar}. In the previous study of the Coulomb interaction effects on
the tilted Dirac fermions\cite{YWLEE}, we have employed a regularization scheme in which the modes with
large energies, not the large momenta, are integrated out. For type-I Dirac fermions, this method yields
the same results as those by integrating out the modes with large momenta. This is because the Fermi
surface is point-like so that large momenta imply large energies. For type-II Dirac fermions, however,
this scheme is necessary since the Fermi surface becomes extended.

For $2$D Dirac fermions, there are three types of disorder: the random scalar potential (RSP), the
random vector potentials along and perpendicular to the tilting direction (referred to as the $x$-RVP
and $y$-RVP, respectively), and the random mass (RM)\cite{Ludwig}. For type-I DSMs, the effects of all
three types of disorder have been examined in Ref. \onlinecite{ZKYang} by an RG analysis of a replica
field theory. Although we have performed the RG transformations on different objects and adopted
different regularization schemes (and thus the resulting RG equations may be distinct), the RG flows of
various types of disorder strengths in type-I DSMs are similar. However, the interpertation of the
resulting ground state is distinct in some situations (see below). On the other hand, the effects of
quenched disorder on type-II DSMs has not been studied before. Our main findings are summarized in Figs.
\ref{wldff2}, \ref{wldff21}, \ref{wldff22}, and Table \ref{T1}. We describe them briefly in the
following.

(i) For the weak RSP or $x$-RVP, the fermion-disorder coupling flows to strong disorder strength at low
energies for both types of Dirac fermions. We assert that the corresponding ground states are insulating
for both cases. The phase diagram is shown in Fig. \ref{wldff2}. For type-I DSMs, our result is in
contrast with the previous work\cite{ZKYang}, where it was claimed, based on the analysis of the fermion
spectrum of the kinetic energy part of the renormalized Hamiltonian, that the ground state should be a
DM with a bulk Fermi arc. In our opinion, this claim can be made only when the fermion-disorder strength
is marginal or irrelevant.

(ii) For the weak $y$-RVP, the fermion-disorder coupling in type-I Dirac fermions is marginal and the
resulting ground state is a semimetal (SM) with dynamical critical exponent $z>1$. These results are
consistent with those of Ref. \onlinecite{ZKYang}. However, we further perform a replica mean-field
analysis to study the effects of the marginal fermion-disorder coupling, which shows that there is a
critical disorder strength, beyond which, we obtain a solution corresponding to the DM. It follows from
the general consideration on the fluctuation effects around the mean-field solution, which are described
by a $2$D generalized nonlinear $\sigma$ model\cite{Hikami}, we assert that this DM is unstable toward
an insulating state. Thus, there are two phases for the type-I DSM: the SM at weak disorder and the
insulating phase beyond the critical disorder strength. Since the critical disorder strength is a
decreasing function of the tilting angle, the SM is, in fact, fragile at moderate magnitude of the
tilting angle.

For type-II Dirac fermions, the fermion-disorder coupling flows to strong disorder strength at low
energies. Hence, we expect that the resulting ground state is insulating. By combining these results,
the phase diagram in the presence of weak $y$-RVP is shown in Fig. \ref{wldff21}.

(iii) Finally, for the weak RM, the fermion-disorder coupling in the type-I DSM is marginally irrelevant.
Moreover, the effective tilt is suppressed at low energies so that the ground state is an untilted DSM.
These are identical to the results of Ref. \onlinecite{ZKYang}.

On the other hand, for type-II Dirac fermions, the fermion-disorder coupling is marginal and the
dynamical critical exponent $z>1$. By calculating the fermion self-energy within the first-order Born
approximation, we find that the quasiparticles acquire a non-zero mean-free time. This suggests that
this state is a DM. Based on the conventional wisdom\cite{Hikami,Abrahams}, this DM is unstable in the
presence of arbitrarily weak disorder and the ground state is insulating.

Since in the presence of weak RM, the ground state of type-I Dirac fermions is an untilted DSM and
the type-II Dirac fermion is in an insulating phase, we expect that there is a DSM-Insulator transition
upon varying the tilting angle for a fixed disorder strength. The schematic phase diagram at weak
disorder is shown in Fig. \ref{wldff22}.

The organization of the rest of the paper is as follows. The model is defined and discussed in Sec.
\ref{model}. We present the one-loop RG equations and its implications in Secs. \ref{rsp}, \ref{rvp},
and \ref{rm} for the RSP (and $x$-RVP), $y$-RVP, and RM, respectively. The last section is devoted to
conclusive discussions. The derivation of the one-loop RG equations are put in the appendix.

\section{The model}
\label{model}

We first introduce the minimal model of a disordered tilted DSM whose Hamiltonian is given by
$H=H_0+H_{\mbox{dis}}$ where
\begin{equation}
 H_0=\sum_{\xi,\sigma,\bm{p}}\tilde{\psi}^{\dagger}_{\xi\sigma}(\bm{p})(\xi wv_1p_1+\xi v_1p_1\sigma_1
 +v_2p_2\sigma_2)\tilde{\psi}_{\xi\sigma}(\bm{p}) \ , \label{wf2h1}
\end{equation}
describes a non-interacting tilted DSM\cite{MOGoerbig}. Here $\xi=\pm 1$ denote the valley degeneracy,
$\sigma=\pm 1$ account for the spin degeneracy, and $\sigma_{1,2,3}$ are the Pauli matrices describing
the conduction-valence band degrees of freedom. The fermionic fields $\tilde{\psi}_{\xi\sigma}(\bm{p})$
and $\tilde{\psi}^{\dagger}_{\xi\sigma}(\bm{p})$ obey the canonical anticommutation relations. Without
loss of generality, we take the velocities $v_1,v_2>0$. The dimensionless quantity $w$ is called the
tilting parameter. The Dirac cone is tilted along the $x$-axis when $w\neq 0$. $|w|<1$ and $|w|>1$
correspond to the type-I and type-II DSM, respectively. We notice that $H_0$ is invariant against the
particle-hole (PH) transformation
\begin{equation}
 \tilde{\psi}_{\xi\sigma}(\bm{p})\rightarrow\sigma_1\tilde{\psi}^*_{\xi\sigma}(-\bm{p}) \ ,
 \label{wf2ph1}
\end{equation}
when the chemical potential $\mu=0$. This PH symmetry forbids terms like
$\tilde{\psi}_{\xi\sigma}^{\dagger}\tilde{\psi}_{\xi\sigma}$ or
$\tilde{\psi}_{\xi\sigma}^{\dagger}\sigma_{1/2}\tilde{\psi}_{\xi\sigma}$.

The spectrum of $H_0$ is
\begin{equation}
 \epsilon_{\pm}(\bm{p})=\xi wv_1p_1\pm\sqrt{v_1^2p_1^2+v_2^2p_2^2} \ , \label{wf2h11}
\end{equation}
for each valley. Here we have set the energy of the Dirac point to be zero. When $\mu=0$, the Fermi
surface for type-I Dirac fermions consists of a single point for each valley, while it consists of two
straight lines:
\begin{equation}
 \tilde{p}_2=\pm\tilde{w}\tilde{p}_1 \ , \label{wf2fs1}
\end{equation}
for type-II Dirac fermions, where $\tilde{p}_a=v_ap_a$ with $a=1,2$ and $\tilde{w}=\sqrt{w^2-1}$. One
may regard each line as a branch of the Fermi surface, and thus the $+$ and $-$ signs are the labels of
the branches. The Fermi-surface topology changes from $|w|<1$ to $|w|>1$. $|w|=1$ corresponds to the
Lifshitz transition point at which the Fermi surface reduces to a single line, given by $p_2=0$ for the
present model.

The Hamiltonian $H_{\mbox{dis}}$, describing the coupling between the Dirac fermions and a random field
$A(\bm{r})$, is of the form
\begin{equation}
 H_{\mbox{dis}}=-\! \sum_{\xi,\sigma} \! \! \int \!d^2x\psi_{\xi\sigma}^{\dagger}\Gamma\psi_{\xi\sigma}
 A(\bm{r}) \ , \label{wf2h12}
\end{equation}
where $\psi_{\xi\sigma}(\bm{r})$ is the inverse Fourier transform of $\tilde{\psi}_{\xi\sigma}(\bm{p})$.
The random field $A(\bm{r})$ is nonuniform and random in space, but constant in time. Thus, it mixes up
the momenta but not the frequencies. We further assume that it is a quenched, Gaussian white-noise field
with the correlation functions:
\begin{equation}
 \langle A(\bm{r})\rangle=0 \ , ~~
 \langle A(\bm{r}_1)A(\bm{r}_2)\rangle=\Delta\delta(\bm{r}_1-\bm{r}_2) \ . \label{wf2dis1}
\end{equation}
and the variance $\Delta$ is chosen to be dimensionless.

In two dimensions, there are three types of disorder\cite{Ludwig}, corresponding to $\Gamma=u_0\sigma_0$,
$\Gamma=u_{1,2}\sigma_{1,2}$, and $\Gamma=u_3\sigma_3$, provided that the random field does not mix the
Dirac fermions with different spins and valley indices, where $\sigma_0$ is the $2\times 2$ unit matrix
and $u_i$ with $i=0,1,2,3$ measures the strength of the single-impurity potential of the corresponding
type of disorder. Since we have chosen $\Delta$ to be dimensionless, $u_i$ has the dimension of speed.
$\Gamma=u_0\sigma_0$, $\Gamma=u_1\sigma_1$, $\Gamma=u_2\sigma_2$, and $\Gamma=u_3\sigma_3$ describe the
RSP, the $x$-RVP, the $y$-RVP, and the RM, respectively. For the $2$D materials like graphene, the RSP
can be produced by adsorbed atoms and vacancies, the RVP comes from the spatial distortion of the $2$D
sheet by ripples\cite{CastroNeto,Peres,Mucciolo} and the RM can be introdcued by the underlying
substrate\cite{Champel}. Although the RSP and RVP break the PH symmetry for a given impurity
configuration, they preserve this symmetry on average.

We will see later that within our model, the RSP and the $x$-RVP will mix at the one-loop order as long
as $w\neq 0$. (That is, the RSP and the RVP in the tilting direction will generate each other under the
RG transformations.) Thus, the two types of disorder must be considered together. On the other hand, the
$y$-RVP and the RM can exist on its own without generating other types of disorder. Therefore, we will
study the effects of each of them separately.

The other effect arising from a non-zero tilting parameter $w$ is that the term
$\psi^{\dagger}_{\xi\sigma}\sigma_1\partial_{\tau}\psi_{\xi\sigma}$ will be generated\cite{Sikkenk,ZKYang}.
Thus, the working action $S$ in the imaginary-time formulation can be written as
\begin{equation}
 S=\! \sum_{\xi,\sigma} \! \! \int \! d\tau d^2x(\mathcal{L}_0+\mathcal{L}_i) \ , \label{wldfs12}
\end{equation}
where
\begin{equation}
 \mathcal{L}_0=\psi^{\dagger}_{\xi\sigma}[(1+\lambda\sigma_1)\partial_{\tau}-i\xi v_1(w+\sigma_1)
 \partial_1-iv_2\sigma_2\partial_2]\psi_{\xi\sigma} \ , \label{wldfs1}
\end{equation}
describes the non-interacting tilted Dirac fermions and $\mathcal{L}_i$ is the coupling to the random
field. For the RSP or $x$-RVP
\begin{equation}
 \mathcal{L}_i=-\psi_{\xi\sigma}^{\dagger}\Gamma\psi_{\xi\sigma}A(\bm{r}) \ , \label{wldfs11}
\end{equation}
with $\Gamma=u_0\sigma_0+u_1\sigma_1$, and
\begin{equation}
 \mathcal{L}_i=-u_j\psi_{\xi\sigma}^{\dagger}\sigma_j\psi_{\xi\sigma}A(\bm{r}) \ , \label{wldfs13}
\end{equation}
for the $y$-RVP ($j=2$) and the RM ($j=3$).

In the following, we would like to study the effects of $\mathcal{L}_i$ on the system with the help of
the RG. Instead of integrating out the random field $A$ to obtain a replica field theory, we will
integrate out the disorder at each order in the perturbation theory. This provides us some technical
advantages.

To properly perform the RG transformations such that they scale toward the Fermi surface, we parametrize
the low-energy degrees of freedom by their energies and an additional dimensionless parameter. Given an
energy $E$, the equal-energy curve is $\epsilon_{\pm}(\bm{p})=E$. For type-I Dirac fermions ($|w|<1$),
this equal-energy curve is an ellipse and can be parametrized as
\begin{eqnarray}
 \tilde{p}_1 &=& -\frac{\xi w}{1-w^2}E+\frac{|E|}{1-w^2}\cos{\theta} \ , \nonumber \\
 \tilde{p}_2 &=& \frac{|E|}{\sqrt{1-w^2}}\sin{\theta} \ , \label{wf2e1}
\end{eqnarray}
where $0\leq\theta<2\pi$. On the other hand, for type-II Dirac fermions ($|w|>1$), this equal-energy
curve is a hyperbola and can be parametrized as
\begin{eqnarray}
 \tilde{p}_1 &=& \frac{\xi w}{w^2-1}E\pm\frac{|E|}{w^2-1}\cosh{\theta} \ , \nonumber \\
 \tilde{p}_2 &=& \frac{|E|}{\sqrt{w^2-1}}\sinh{\theta} \ , \label{wf2e11}
\end{eqnarray}
where $-\infty<\theta<+\infty$. The $+$ and $-$ signs correspond to the right and the left branches of
the hyperbola, respectively.

To proceed, we separate the fermion fields $\psi_{\xi\sigma}$ into the slow and fast modes. The slow
modes $\psi_{\xi\sigma<}$ and the fast modes $\psi_{\xi\sigma>}$ contain excitations in the energy range
$|E|<\Lambda/s$ and the energy shell $\Lambda/s<|E|<\Lambda$, respectively, where $\Lambda$ is the UV
cutoff in energies and $s=e^l>1$. By integrating out the fast modes of fermion fields to the one-loop
order, we obtain an effective action of the slow modes. We then rescale the variables and fields by
\begin{eqnarray}
 & & E\rightarrow e^{-l}E \ , ~\theta\rightarrow\theta \ , ~\tau\rightarrow e^{zl}\tau \ , \nonumber \\
 & & \psi_{\xi\sigma<}\rightarrow Z_{\psi}^{-1/2}\psi_{\xi\sigma} \ , ~A\rightarrow e^{-l}A \ ,
     \label{wf2rg12}
\end{eqnarray}
to bring the term $\psi^{\dagger}_{\xi\sigma<}\partial_{\tau}\psi_{\xi\sigma<}$ in the action back to the
original form. In this way, we obtain a set of one-loop RG equations for the parameters in the action $S$.
We will list the one-loop RG equations for each type of disorder in the following sections, and leave the
details of their derivation to the appendix.

\section{The RSP and $x$-RVP}
\label{rsp}
\subsection{Type-I DSMs}

We first consider the RSP and $x$-RVP. For type-I Dirac fermions, the renormalized parameters are given
by
\begin{eqnarray*}
 w^{\prime} \! \! &=& \! \! w \ , \\
 \frac{v_{1,2}^{\prime}}{v_{1,2}} \! \! &=& \! \! 1+ \! \left[z-1
 -\frac{\Delta(1-w\lambda)(u_0^2+u_1^2-2wu_0u_1)}{2\pi v_1v_2(1-w^2)^{3/2}}\right] \! l \\
 \! \! & & \! \! +O(l^2) \ , \\
 \lambda^{\prime} \! \! &=& \! \! \lambda
 -\frac{\Delta[(w+\lambda)(u_0^2+u_1^2)-2(1+w\lambda)u_0u_1]}{2\pi v_1v_2(1-w^2)^{3/2}} \\
 & & \times(1-w\lambda)l+O(l^2) \ , \\
 u_0^{\prime} \! \! &=& \! \! u_0+\! \left[z-1-\frac{\Delta(1-w\lambda)(u_0^2+u_1^2-2wu_0u_1)}
 {2\pi v_1v_2(1-w^2)^{3/2}}\right] \! u_0l \\
 \! \! & & \! \! +\frac{\Delta[u_0^3-wu_1^3-3wu_0^2u_1+(1+2w^2)u_1^2u_0]}{2\pi v_1v_2(1-w^2)^{3/2}}l \\
 \! \! & & \! \! +O(l^2) \ , \\
 u_1^{\prime} \! \! &=& \! \! u_1+\! \left[z-1-\frac{\Delta(1-w\lambda)(u_0^2+u_1^2-2wu_0u_1)}
 {2\pi v_1v_2(1-w^2)^{3/2}}\right] \! u_1l \\
 \! \! & & \! \! -\frac{\Delta [wu_0^3-w^2u_1^3+3wu_1^2u_0-(2+w^2)u_0^2u_1]}{2\pi v_1v_2(1-w^2)^{3/2}}l
 \\
 \! \! & & \! \! +O(l^2) \ .
\end{eqnarray*}
If we choose $v_{1,2}$ to be RG invariants, then we have
\begin{equation}
 z=1+\frac{\Delta(1-w\lambda)(u_0^2+u_1^2-2wu_0u_1)}{2\pi v_1v_2(1-w^2)^{3/2}} \ , \label{wf2rg14}
\end{equation}
which leads to
\begin{eqnarray*}
 \lambda^{\prime} &=& \lambda-\frac{\Delta[(w+\lambda)(u_0^2+u_1^2)-2(1+w\lambda)u_0u_1]}
 {2\pi v^2(1-w^2)^{3/2}} \\
 & & \times(1-w\lambda)l+O(l^2) \ , \\
 u_0^{\prime} &=& u_0+\frac{\Delta[u_0^3-wu_1^3-3wu_0^2u_1+(1+2w^2)u_1^2u_0]}{2\pi v^2(1-w^2)^{3/2}}l \\
 & & +O(l^2) \ , \\
 u_1^{\prime} &=& u_1-\frac{\Delta [wu_0^3-w^2u_1^3+3wu_1^2u_0-(2+w^2)u_0^2u_1]}{2\pi v^2(1-w^2)^{3/2}}l
 \\
 & & +O(l^2) \ .
\end{eqnarray*}
For simplicity, we have set $v_1=v_2=v$. Consequently, we get the one-loop RG equations for $\lambda$,
$u_1$, and $u_2$
\begin{eqnarray}
 \frac{d\lambda_l}{dl} &=&
 [2(1+w\lambda_l)\gamma_{0l}\gamma_{1l}-(w+\lambda_l)(\gamma_{0l}^2+\gamma_{1l}^2)] \nonumber \\
 & & \times(1-w\lambda_l) \ , \label{wf2rge17}
\end{eqnarray}
and
\begin{eqnarray}
 \frac{d\gamma_{0l}}{dl} \! \! &=& \! \! (\gamma_{0l}-w\gamma_{1l})
 (\gamma_{0l}^2-2w\gamma_{0l}\gamma_{1l}+\gamma_{1l}^2) \ , \label{wf2rge13} \\
 \frac{d\gamma_{1l}}{dl} \! \! &=& \! \! -(\gamma_{0l}-w\gamma_{1l})
 (w\gamma_{0l}^2-2\gamma_{0l}\gamma_{1l}+w\gamma_{1l}^2) \ , ~~~~~~\label{wf2rge14}
\end{eqnarray}
where the quantities with subscript $l$ refer to those at the scale $l$, the ones without the subscript
refer to the bare values ($l=0$), and
\begin{eqnarray*}
 \gamma_{0(1)l}=\sqrt{\frac{\Delta}{2\pi(1-w^2)^{3/2}}}\frac{u_{0(1)l}}{v}
\end{eqnarray*}
are the dimensionless fermion-disorder couplings.

\begin{figure}
\begin{center}
 \includegraphics[width=0.99\columnwidth]{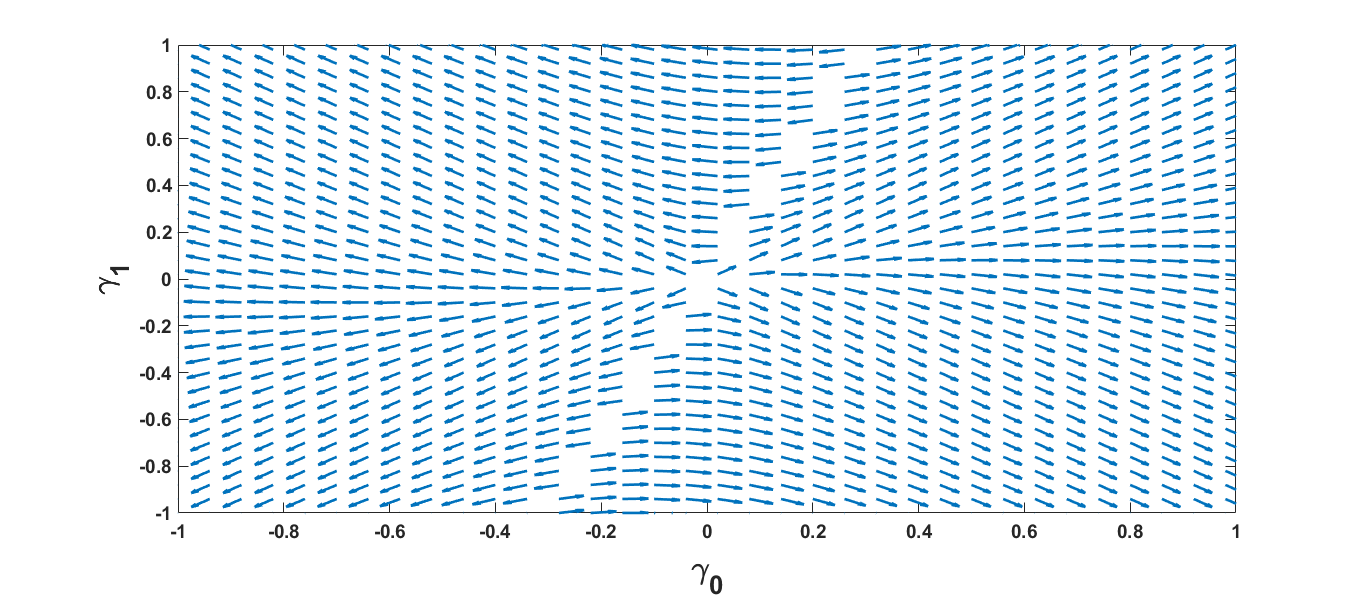}
 \caption{The RG flow of $\gamma_{0l}$ and $\gamma_{1l}$ for type-I DSMs with $w=0.3$. The fixed line
 $\gamma_0=w\gamma_1$ is IR unstable.}
 \label{wldff11}
\end{center}
\end{figure}

The typical RG flow of $\gamma_{0l}$ and $\gamma_{1l}$ is depicted in Fig. \ref{wldff11}. Equations
(\ref{wf2rge13}) and (\ref{wf2rge14}) have a fixed line $\gamma_0=w\gamma_1$. The RSP and $x$-RVP
correspond to the lines with $\gamma_1=0$ and $\gamma_0=0$, respectively. The RG flow for the RSP and
$x$-RVP are described in the following\cite{foot1}.

We first consider the RSP, i.e., $\gamma_1=0$. If we start with $\gamma_0>0$ and $\gamma_1=0$, then for
$w>0$, $\gamma_{0l}$ will increase and $\gamma_{1l}$ will decrease with increasing $l$. Hence, at low
energies, we get $\gamma_{0l}\rightarrow +\infty$ and $\gamma_{1l}\rightarrow -\infty$. On the other
hand, if we start with $\gamma_0<0$ and $\gamma_1=0$, then for $w>0$, $\gamma_{0l}$ will decrease and
$\gamma_{1l}$ will increase with increasing $l$. Hence, at low energies, we get
$\gamma_{0l}\rightarrow -\infty$ and $\gamma_{1l}\rightarrow +\infty$. This means that the type-I DSM is
unstable in the presence of weak RSP. Since the disorder strength becomes strong at low energies, we
expect that the resulting ground state is insulating.

Next, we consider the $x$-RVP, i.e., $\gamma_0=0$. If we start with $\gamma_0=0$ and $\gamma_1>0$, then
for $w>0$, $\gamma_{0l}$ will decrease and $\gamma_{1l}$ will increase with increasing $l$. Hence, at
low energies, we get $\gamma_{0l}\rightarrow -\infty$ and $\gamma_{1l}\rightarrow +\infty$. On the other
hand, if we start with $\gamma_0=0$ and $\gamma_1<0$, then for $w>0$, $\gamma_{0l}$ will increase and
$\gamma_{1l}$ will decrease with increasing $l$. Hence, at low energies, we get
$\gamma_{0l}\rightarrow +\infty$ and $\gamma_{1l}\rightarrow -\infty$. This means that the type-I DSM is
unstable in the presence of weak $x$-RVP. Since the disorder strength becomes strong at low energies, we
expect that the resulting ground state is also insulating.

\begin{figure}
\begin{center}
 \includegraphics[width=0.99\columnwidth]{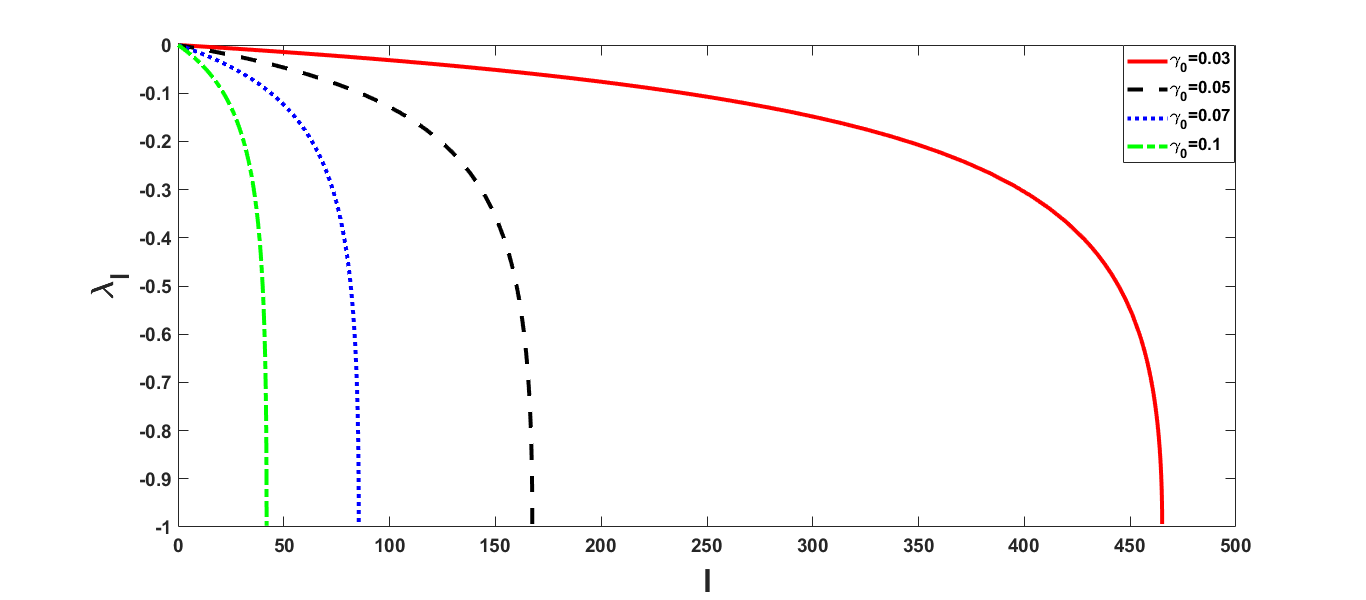}
 \caption{The RG flow of $\lambda_l$ for a type-I DSM in the presence of RSP ($\gamma_1=0=\lambda$) with
 $w=0.3$ and various bare values of $\gamma_{0l}$. The flow of $\lambda_l$ stops when one of $\gamma_{0l}$
 and $\gamma_{1l}$ becomes divergent.}
 \label{wldff14}
\end{center}
\end{figure}

The RG flow of $\lambda_l$ in the presence of the RSP, with various values of $\gamma_0$, is shown in
Fig. \ref{wldff14}. We see that $\lambda_l\rightarrow -\eta_w$ at some critical value $l_c$ where one of
$\gamma_{0l}$ and $\gamma_{1l}$ becomes divergent. For given $w$, the value of $l_c$ decreases with the
increasing value of $\gamma_0$. The situation is similar for the $x$-RVP.

Although our RG scheme is different from that adopted in Ref. \onlinecite{ZKYang}, the RG flows of
the fermion-disorder couplings $\gamma_{0l}$, $\gamma_{1l}$ and the parameter $\lambda_l$ are similar.
However, our interpretation of the resulting physics is distinct from that in Ref. \onlinecite{ZKYang}.
There, the authors consider only the kinetic energy part of the renormalized Hamiltonian and claim that
the resulting phase is a DM with a bulk Fermi arc. In our opinion, this is justified only when the
fermion-disorder couplings are marginal or irrelevant. Then, they can be regarded as perturbations and
the kinetic energy part of the renormalized Hamiltonian dominates the low-energy physics. In the present
case, the fermion-disorder couplings are relevant operators, exhibiting runaway RG flows, so that the
low-energy physics is dominated by these terms. When the disorder potential becomes strong, we expect
that the electrons are localized at the minia of the potential and the system is an insulator.

\subsection{Type-II DSMs}

\begin{figure}
\begin{center}
 \includegraphics[width=0.99\columnwidth]{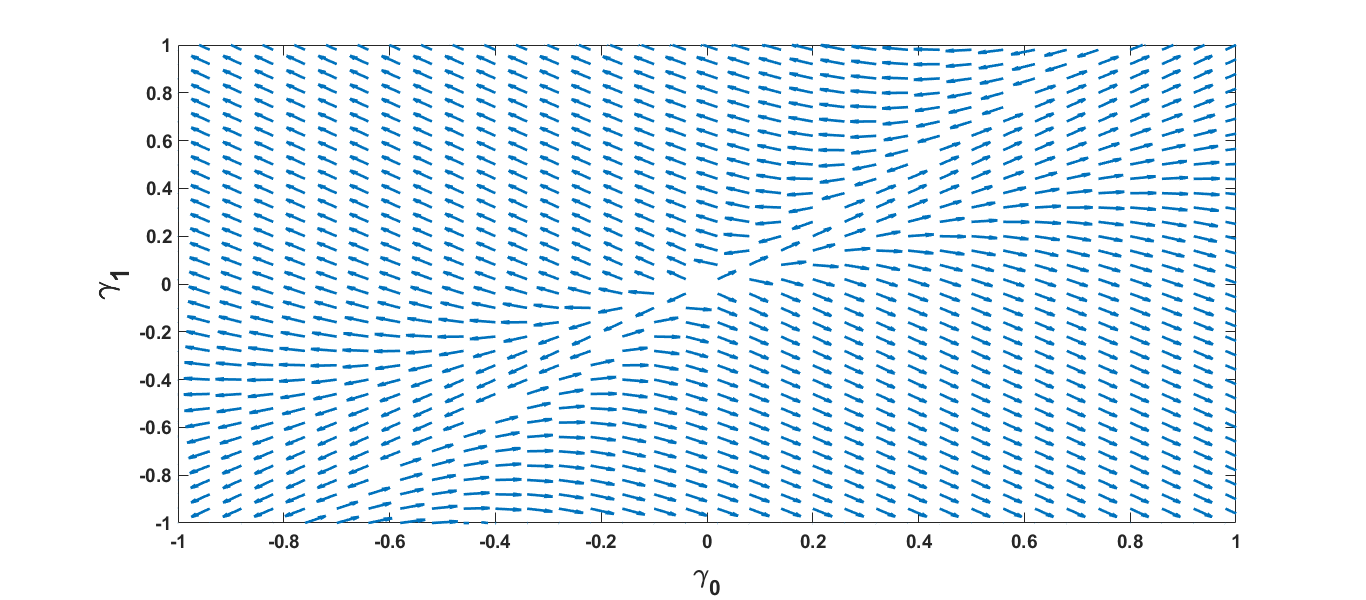}
 \caption{The RG flow of $\gamma_{0l}$ and $\gamma_{1l}$ for type-II DSMs with $w=1.3$. Notice the fixed
 line at $w\gamma_0=\gamma_1$.}
 \label{wldff12}
\end{center}
\end{figure}

Next, we consider the type-II DSMs. Similar to type-I DSMs, we find that $w^{\prime}=w$ to the one-loop
order. Moreover, we choose the value of $z$ to be
\begin{equation}
 z=1+\frac{\Delta(w-\lambda)[w(u_0^2+u_1^2)-2u_0u_1]}{2\pi^2v_1v_2|w|(w^2-1)} \ , \label{wf2rgsol2}
\end{equation}
so that both $v_1$ and $v_2$ are RG invariants. Thus, we may set $v_1=v_2=v$ for simplicity. Accordingly,
the one-loop RG equations for $\lambda$, $u_1$, and $u_2$ are
\begin{equation}
 \frac{d\lambda_l}{dl}=(\lambda_l-w)[(1+w\lambda_l)(\gamma_{0l}^2+\gamma_{1l}^2)-2(w+\lambda_l)
 \gamma_{0l}\gamma_{1l}] \ , \label{wf2rge2}
\end{equation}
and
\begin{eqnarray}
 \frac{d\gamma_{0l}}{dl} &=& (w\gamma_{0l}-\gamma_{1l})(w\gamma_{0l}^2-2\gamma_{0l}\gamma_{1l}
 +w\gamma_{1l}^2) \ , ~~~~~~\label{wf2rge21} \\
 \frac{d\gamma_{1l}}{dl} &=& -(w\gamma_{0l}-\gamma_{1l})(\gamma_{0l}^2-2w\gamma_{0l}\gamma_{1l}
 +\gamma_{1l}^2) \ , \label{wf2rge22}
\end{eqnarray}
where
\begin{eqnarray*}
 \gamma_{0((1)l}=\sqrt{\frac{\Delta}{2\pi^2|w|(w^2-1)}}\frac{u_{0(1)l}}{v} \ .
\end{eqnarray*}
Equations (\ref{wf2rge21}) and (\ref{wf2rge22}) have a fixed line $w\gamma_0=\gamma_1$. The typical RG
flow of $\gamma_{0l}$ and $\gamma_{1l}$ is depicted in Fig. \ref{wldff12}. We notice that the qualitative
behaviors of the RG flow for $\gamma_{0l}$ and $\gamma_{1l}$ are similar for both types of DSMs. As a
result, similar to type-I DSMs, the ground state is an insulator for type-II DSMs in the presence of
weak RSP or $x$-RVP.

\begin{figure}
\begin{center}
 \includegraphics[width=0.99\columnwidth]{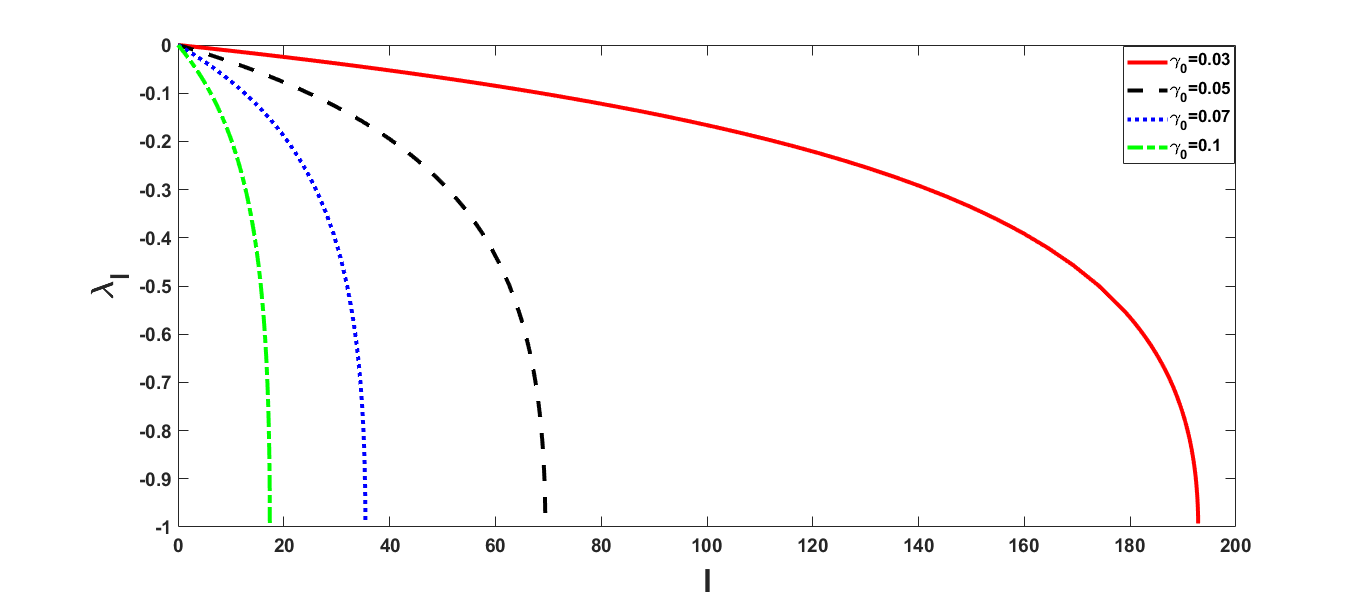}
 \caption{The RG flow of $\lambda_l$ for a type-II DSM in the presence of RSP ($\gamma_1=0=\lambda$) with
 $w=1.3$ and various bare values of $\gamma_{0l}$. The flow of $\lambda_l$ stops when one of $\gamma_{0l}$
 and $\gamma_{1l}$ becomes divergent.}
 \label{wldff15}
\end{center}
\end{figure}

The RG flow of $\lambda_l$ in the presence of the RSP, with various values of $\gamma_0$, is shown in
Fig. \ref{wldff15}. The qualitative behavior is similar to type-I DSMs. $\lambda_l\rightarrow -\eta_w$
at some critical value $l_c$ where one of $\gamma_{0l}$ and $\gamma_{1l}$ becomes divergent. For given
$w$, the value of $l_c$ decreases with the increasing value of $\gamma_0$. The case with the $x$-RVP is
similar. The only difference between type-I and type-II DSMs is that $l_c$ is smaller for the latter
with the same value of $\gamma_0$.

\begin{figure}
\begin{center}
 \includegraphics[width=0.9\columnwidth]{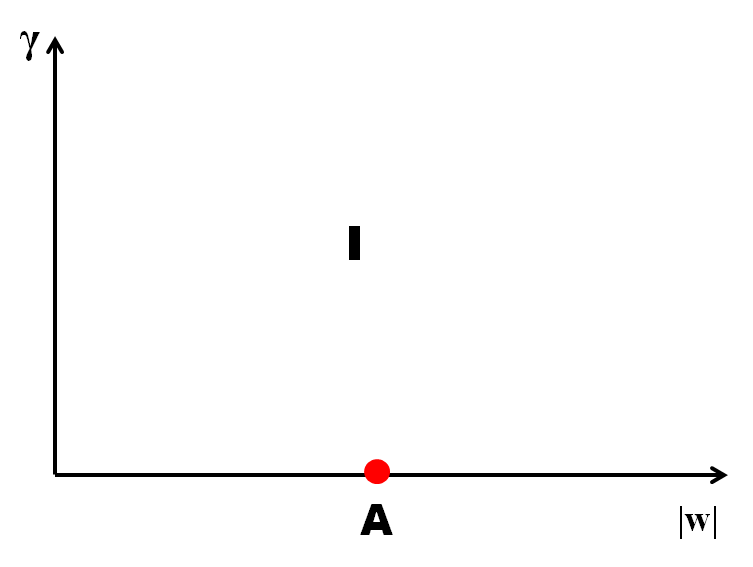}
 \caption{The phase diagram of a non-interacting tilted DSM in the presence of RSP or $x$-RVP. $\gamma$
 and $w$ are the (dimensionless) disorder strength and the tilting parameter, respectively. I denotes
 the insulator. Point A located at $|w|=1$ and $\gamma=0$ is the Lifshitz transition point, separating
 the type-I and type-II DSMs in the absence of disorder.}
 \label{wldff2}
\end{center}
\end{figure}

From the above analysis, we expect that the behaviors of the system at finite disorder strength in the
regions with $|w|<1$ and $|w|>1$ are qualitatively similar to each other. That is, in the presence of
the RSP or $x$-RVP, there is no phase transition from $|w|<1$ to $|w|>1$, and the ground state is an
insulator. A schematic phase diagram in the presence of the RSP or $x$-RVP is shown in Fig. \ref{wldff2}.

\section{The $y$-RVP}
\label{rvp}
\subsection{Type-I DSMs}

Next, we consider the $y$-RVP. For type-I DSMs, we find that $w^{\prime}=w$ to the one-loop order. If we
choose
\begin{equation}
 z=1+\frac{\Delta u_2^2}{2\pi v_1v_2(1-w^2)^{3/2}}(1-w\lambda) \ , \label{wf2rg15}
\end{equation}
then $v_1$, $v_2$, and $u_2$ are all marginal at the one-loop order. On the other hand, the one-loop RG
equation for $\lambda$ is
\begin{equation}
 \frac{d\lambda_l}{dl}=\frac{\Delta u_2^2}{2\pi v_1v_2(1-w^2)^{3/2}}(1-w\lambda_l)(w-\lambda_l) \ .
 \label{wf2rge18}
\end{equation}

The solution of Eq. (\ref{wf2rge18}) with the initial value $\lambda=0$ is given by
\begin{equation}
 \frac{\lambda_l-w}{\lambda_l-1/w}=w^2
 \exp{\! \left[-\frac{\Delta u_2^2l}{2\pi v_1v_2\sqrt{1-w^2}}\right]} . \label{wf2rgsol1}
\end{equation}
At low energies, i.e., $l\rightarrow +\infty$, we get $\lambda_*=\lambda_{+\infty}=w$\cite{foot2}.
Inserting this value of $\lambda_l$ into Eq. (\ref{wf2rg15}), we get a non-universal dynamical exponent
$z$ given by $z=1+\eta$, where
\begin{equation}
\eta=\frac{\Delta u_2^2}{2\pi v_1v_2\sqrt{1-w^2}} \ . \label{wf2rg17}
\end{equation}
A non-zero value of $\lambda_*$ will affect the dispersion relation of quasiparticles, which is
determined by the poles of the single-particle propagator on the complex frequency plane with the
replacement
\begin{eqnarray*}
 v_{1(2)}\rightarrow v_{1(2)} \! \left[\frac{p}{k_0}\right]^{\! \eta} ,
\end{eqnarray*}
where $p=|\bm{p}|$, $k_0\sim 1/a_0$, and $a_0$ is the lattce spacing. As a result, the dispersion
relation of the quasiparticles is given by
\begin{equation}
 E_{\pm}(\bm{p})=\pm \! \left(\frac{p}{k_0}\right)^{\! \eta} \!
 \sqrt{v_1^2p_1^2+\frac{v_2^2}{1-w^2}p_2^2} \ . \label{wf2rgsol11}
\end{equation}
To sum up, in the presence of weak $y$-RVP, the system is a SM with $z>1$ and the fermion-disorder
coupling is marginal.

For a SM, various physical quantities will exhibit power-law temperature dependence at low temperatures.
This has been discussed in Ref. \onlinecite{ZKYang}, and we will not duplicate it here. Instead, we ask
the question. Will this SM be stable against the presence of a marginal fermion-disorder coupling? As
is well known, the disorder potential is a marginal perturbation to the FL, and a $2$D FL is unstable
toward an insulator even in the presence of an arbitrarily weak disorder potential\cite{Abrahams}. To
answer this question, we employ the replica trick to map the renormalized action into a replica field
theory and then perform a mean-field analysis.

The disorder-averaged replicated partition function $\mathcal{Z}$ in the imaginary-time formulation is
given by\cite{Finkelshtein,Belitz,Altland}
\begin{eqnarray*}
 \mathcal{Z}=\! \int \! D[A]P[A] \! \int \! D[\chi]D[\bar{\chi}]e^{-S_0-S_i} \ ,
\end{eqnarray*}
where
\begin{eqnarray*}
 S_0 &=& \! \sum_{a=1}^M \! \sum_{\xi,n} \! \int \! d^2x\bar{\chi}_{\xi na}[-i\omega_n(1+w\sigma_1)
 +\hat{h}_{\xi}]\chi_{\xi na} \\
 &=& \! \sum_{\xi,n} \! \int \! d^2x\bar{\psi}_{\xi n}[-i\omega_n(1+w\sigma_1)+\hat{h}_{\xi}]\otimes I_M
 \psi_{\xi n} \ , \\
 S_i &=& -u_2 \! \sum_{a=1}^M \! \sum_{\xi,n} \! \int \! d^2x\bar{\chi}_{\xi na}\sigma_2\chi_{\xi na}A
 \\
 &=& -u_2 \! \sum_{\xi,n} \! \int \! d^2x\bar{\psi}_{\xi n}\sigma_2\otimes I_M\psi_{\xi n}A \ ,
\end{eqnarray*}
and
\begin{eqnarray*}
 P[A]=\exp{\! \left(-\frac{1}{2\Delta} \! \int \! d^2xA^2\right)} .
\end{eqnarray*}
In the above, $\omega_n=(2n+1)\pi T$, $a$ is the replica index, $I_M$ is the unit matrix of dimension
$M$ in the replica space, $\psi_{\xi n}=[\chi_{\xi n1},\cdots,\chi_{\xi nM}]^t$,
$\bar{\psi}_{\xi n}=[\bar{\chi}_{\xi n1},\cdots,\bar{\chi}_{\xi nM}]$, and
\begin{eqnarray*}
 \hat{h}_{\xi}=v \! \left(\frac{p}{k_0}\right)^{\! \eta} \! [\xi(w+\sigma_1)p_1+\sigma_2p_2] \ ,
\end{eqnarray*}
in the momentum space. For simplicity, we have set $v_1=v=v_2$ and dropped out the spin index $\sigma$.
By integrating out the random field $A(\bm{r})$, $\mathcal{Z}$ can be written as
\begin{eqnarray*}
 \mathcal{Z}=\! \int \! \! D[\chi]D[\bar{\chi}]\exp{\! \left[-S_0+\frac{g_2^2}{2} \! \! \int \! \! d^2x
 (\bar{\Psi}\Gamma\Psi)^2\right]} ,
\end{eqnarray*}
where $g_2=\sqrt{\Delta u_2^2}$, $\Psi=[\Psi_1,\Psi_{-1}]^t$, $\bar{\Psi}=[\bar{\Psi}_1,\bar{\Psi}_{-1}]$,
$\Gamma_{mn;\xi\xi^{\prime}}^{ab}=\delta_{mn}\delta_{ab}\delta_{\xi\xi^{\prime}}\sigma_2$, and
\begin{eqnarray*}
 \Psi_{\xi} &=& [\cdots,\psi_{\xi 1},\psi_{\xi 0},\psi_{\xi -1},\cdots]^t \ , \\
 \bar{\Psi}_{\xi} &=& [\cdots,\bar{\psi}_{\xi 1},\bar{\psi}_{\xi 0},\bar{\psi}_{\xi -1},\cdots] \ .
\end{eqnarray*}

To proceed, we make a Hubbard-Stratonovich transformation on the four-fermion coupling arising from the
integration over the random field:
\begin{eqnarray*}
 & & \! \! \exp{\! \left[\frac{g_2^2}{2} \! \! \int \! \! d^2x(\bar{\Psi}\Gamma\Psi)^2\right]} \\
 & & \! \! =\! \int \! \! D[Q]\exp{\! \left\{- \! \int \! \! d^2x \! \left[\frac{1}{2}\mbox{tr}Q^2-ig_2
	 \bar{\Psi}Q\Gamma\Psi\right] \! \right\}} ,
\end{eqnarray*}
where $Q^{\dagger}=Q$. If we put an UV cutoff on the frequencies, i.e., $-R\leq n<R$ or
$|\omega_n|\leq (2R-1)\pi T$, then the symmetry group in the absence of the frequency term is U($2RM$).
Under the U($2RM$) transformation,
\begin{equation}
 \Psi_{\xi}\rightarrow U\Psi_{\xi} \ , ~~\bar{\Psi}_{\xi}\rightarrow\bar{\Psi}_{\xi}U^{\dagger} \ ,
 \label{wldfrft1}
\end{equation}
the $Q$ field transforms as
\begin{equation}
 Q_{\xi\xi^{\prime}}\rightarrow UQ_{\xi\xi^{\prime}}U^{\dagger} \ . \label{wldfrft11}
\end{equation}
By integrating out the fermion fields, $\mathcal{Z}$ can be written as
\begin{equation}
 \mathcal{Z}=\! \int \! \! D[Q]e^{-I[Q]} \ , \label{wldfrft12}
\end{equation}
where
\begin{eqnarray}
 I[Q] &=& -\mbox{tr}\ln{\! \left[i\hat{\omega}(1+w\sigma_1)-\hat{h}+ig_2Q\Gamma\right]} \nonumber \\
 & & +\frac{1}{2} \! \int \! \! d^2x\mbox{tr}Q^2 . \label{wldfrft13}
\end{eqnarray}
In Eq. (\ref{wldfrft13}),
$\hat{\omega}_{mn;\xi\xi^{\prime}}^{ab}=\omega_n\delta_{mn}\delta_{ab}\delta_{\xi\xi^{\prime}}$ and
$\hat{h}_{mn;\xi\xi^{\prime}}^{ab}=\hat{h}_{\xi}\delta_{mn}\delta_{ab}\delta_{\xi\xi^{\prime}}$.

We assume that the path integral is dominated by configurations of the $Q$ field close to the homogeneous
solution $Q_0$ of the saddle-point equation $\delta I[Q]/\delta Q=0$. It is given by
\begin{equation}
 Q_0=ig_2 \! \int \! \frac{d^2p}{(2\pi)^2}\mbox{tr}[\Gamma\hat{G}(i\omega_n,\bm{p})] \ ,
 \label{wldfrft14}
\end{equation}
where the trace is taken over the spinor space which describes the conduction-valence band degrees of
freedom and
\begin{equation}
 \hat{G}^{-1}(i\omega_n,\bm{p})=i\hat{\omega}(1+w\sigma_1)-\hat{h}+ig_2Q_0\Gamma \ . \label{wldfrft15}
\end{equation}
To solve Eq. (\ref{wldfrft14}), we try the ansatz
$g_2Q_{\xi\xi^{\prime}}=\alpha_{\xi}\Lambda\delta_{\xi\xi^{\prime}}$, where
$\Lambda_{mn}^{ab}=s_n\delta_{mn}\delta_{ab}$, $s_n=\mbox{sgn}(\omega_n)$, and $\alpha_{\xi}$ is a real
constant which may depend on the valley index $\xi$. Then, $\alpha_{\xi}$ satisfies the equation
\begin{widetext}
\begin{eqnarray*}
 \alpha_{\xi} &=& \! \int_{|p_i|<k_0} \! \frac{d^2p}{(2\pi)^2}
 \frac{2ig_2^2/v^2[i\alpha_{\xi}-s_nv(p/k_0)^{\eta}p_2]}
 {(1-w^2)(p/k_0)^{2\eta}p_1^2+[(p/k_0)^{\eta}p_2-is_n\alpha_{\xi}/v]^2} \\
 &=& \! \int_{|p_i|<k_0} \! \frac{d^2p}{(2\pi)^2}\frac{2ig_2^2/v^2[i\alpha_{\xi}-v(p/k_0)^{\eta}p_2]}
 {(1-w^2)(p/k_0)^{2\eta}p_1^2+[(p/k_0)^{\eta}p_2-i\alpha_{\xi}/v]^2} \ .
\end{eqnarray*}
In the above, we have taken the limit $\omega_n\rightarrow 0$. Moreover, we have changed the variable
$s_np_2\rightarrow s_np_2$. The momentum integral is divergent, and an UV cutoff $k_0$ in momenta is
introduced. We notice that this equation has real solutions. Furthermore, $\alpha_{\xi}$ is independent
of $\xi$, and thus we will set $\alpha_{\xi}\rightarrow\alpha$. Defining the dimensionless quantity
$\tilde{\alpha}=\alpha/(vk_0)$, the above equation becomes
\begin{equation}
 \tilde{\alpha}=\! \int_{\mathcal{D}} \! \frac{d^2x}{(2\pi)^2}
 \frac{2ig_2^2/v^2(i\tilde{\alpha}-r^{\eta}x_2)}{(1-w^2)r^{2\eta}x_1^2+(r^{\eta}x_2-i\tilde{\alpha})^2}
 \ , \label{wldfrft16}
\end{equation}
where $r=\sqrt{x_1^2+x_2^2}$ and $\mathcal{D}=\{(x_1,x_2)||x_1|,|x_2|\leq 1\}$. Equation (\ref{wldfrft16})
has a trivial solution $\tilde{\alpha}=0$. We would like to search for a non-trivial real solution if it
exists. It is clear that if $\tilde{\alpha}$ is a solution of Eq. (\ref{wldfrft16}), then
$-\tilde{\alpha}$ is also a solution. Without loss of generality, we take $\tilde{\alpha}\geq 0$.
	
To find a non-trivial real solution of Eq. (\ref{wldfrft16}), we add the complex conjugate of this
equation to itself, yielding
\begin{eqnarray*}
 \tilde{\alpha}=\! \int_{\mathcal{D}} \! \frac{d^2x}{(2\pi)^2}
 \frac{2g_2^2/v^2\tilde{\alpha}[r^{2\eta}x_2^2+\tilde{\alpha}^2-(1-w^2)r^{2\eta}x_1^2]}
 {[(1-w^2)r^{2\eta}x_1^2+r^{2\eta}x_2^2-\tilde{\alpha}^2]^2+4r^{2\eta}x_2^2\tilde{\alpha}^2} \ .
\end{eqnarray*}
Now the right hand side of this equation becomes real. Therefore, a non-trivial solution satisfies this
equation
\begin{equation}
 1=\frac{4t}{\pi} \! \int^1_0 \! \! dx_1dx_2\frac{r^{2\eta}x_2^2+\tilde{\alpha}^2-(1-w^2)r^{2\eta}x_1^2}
 {[(1-w^2)r^{2\eta}x_1^2+r^{2\eta}x_2^2-\tilde{\alpha}^2]^2+4r^{2\eta}x_2^2\tilde{\alpha}^2} \ ,
 \label{wldfrft17}
\end{equation}
\end{widetext}
where $t=g_2^2/(2\pi v^2)$ measures the disorder strength and $\eta=t/\sqrt{1-w^2}$. Since $vk_0$ can be
regarded as the largest energy scale in this problem, we must have $\tilde{\alpha}<1$.

\begin{figure}
\begin{center}
 \includegraphics[width=0.99\columnwidth]{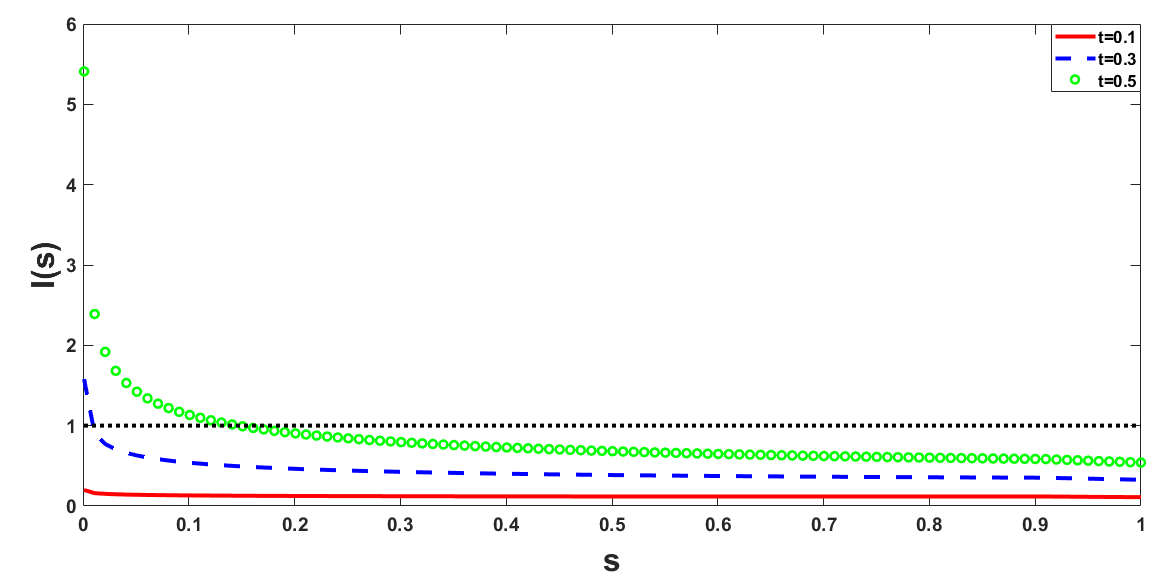}
 \caption{The behavior of $I(s)$ in the range $0\leq s\leq 1$ for different values of $t$ with $|w|=0.3$,
 where $s=\tilde{\alpha}^2$. For reference, $I(s)=1$ is indicated by the dotted line.}
 \label{wldff23}
\end{center}
\end{figure}

We shall solve Eq. (\ref{wldfrft17}) graphically. Let us define the right hand side of Eq.
(\ref{wldfrft17}) as a function of $s=\tilde{\alpha}^2$:
\begin{eqnarray*}
 & & \! \! I(s) \\
 & & \! \! =\! \! \int^1_0 \! \! dx_1dx_2\frac{(4t/\pi)[s+r^{2\eta}x_2^2-(1-w^2)r^{2\eta}x_1^2]}
	 {[(1-w^2)r^{2\eta}x_1^2+r^{2\eta}x_2^2-s]^2+4sr^{2\eta}x_2^2} \ .
\end{eqnarray*}
Figure \ref{wldff23} shows the function $I(s)$ in the range $0\leq s\leq 1$ for different values of $t$
with $|w|=0.3$. We see that for given $|w|$, there exists a critical value $t_c(|w|)$ such that we get a
nontrivial solution of $\tilde{\alpha}$ when $t>t_c(|w|)$. On the other hand, there is only a trivial
solution $\tilde{\alpha}=0$ when $t<t_c(|w|)$. Moreover, for a fixed value of $|w|$, the nontrivial
solution $\tilde{\alpha}$, if it exists, is an increasing function of $t$.

The mean-field solution with $\tilde{\alpha}\neq 0$ has a nonvanishing spectral density at the Fermi
level, and thus corresponds to the DM phase. Since the U($2RM$) symmetry is broken down to
U($RM$)$\times$U($RM$) when $\tilde{\alpha}\neq 0$, there will be Goldstone modes according to the
Goldstone theorem. The DM phase is stable only when it survives the fluctuations of these Goldstone
modes. The latter is described by a certain type of generalized non-linear $\sigma$ models. The RG
analysis of the generalized non-linear $\sigma$ model indicates that the DM phase in $d=2$ is unstable
toward an insulator\cite{Hikami}. On the other hand, the mean-field solution with $\tilde{\alpha}=0$
corresponds to the SM phase with $z>1$. It is stable against small fluctuations around the mean-field
state due to the vanishing DOS at the Fermi level. Hence, we claim that the SM phase with $z>1$ is stable
against the weak $y$-RVP when $t<t_c(|w|)$.

\begin{figure}
\begin{center}
 \includegraphics[width=0.99\columnwidth]{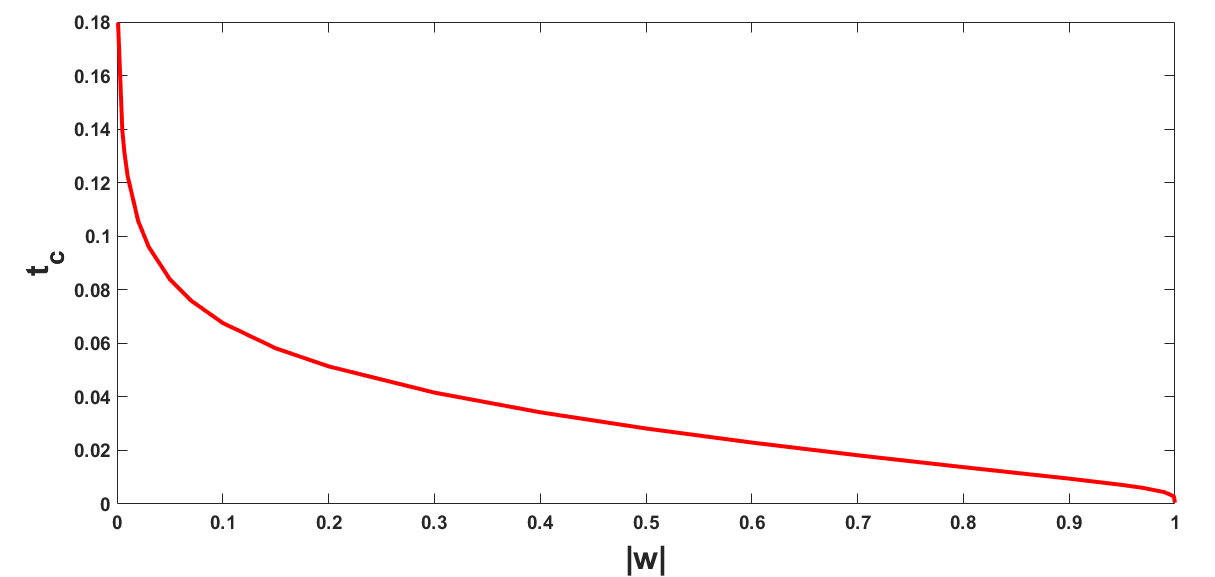}
 \caption{The critical (dimensionless) disorder strength $t_c$ as a function of $|w|$ in the range
 $0\leq |w|<1$.}
 \label{wldff24}
\end{center}
\end{figure}

For given $|w|$, the critical value $t_c(|w|)$ is determined by setting $\tilde{\alpha}=0$ in Eq.
(\ref{wldfrft17}), yielding
\begin{equation}
 1=\frac{4t}{\pi} \! \int^1_0 \! \! dx_1dx_2\frac{r^{2\eta}x_2^2-(1-w^2)r^{2\eta}x_1^2}
 {[(1-w^2)r^{2\eta}x_1^2+r^{2\eta}x_2^2]^2} \ . \label{wldfrft18}
\end{equation}
Equation (\ref{wldfrft18}) can be solved numerically, and the result is shown in Fig. \ref{wldff24}. We
see that $t_c$ is a monotonously decreasing function of $|w|$. Moreover, $t_c\rightarrow 0$ as
$|w|\rightarrow 1$.

\subsection{Type-II DSMs}

\begin{figure}
\begin{center}
 \includegraphics[width=0.9\columnwidth]{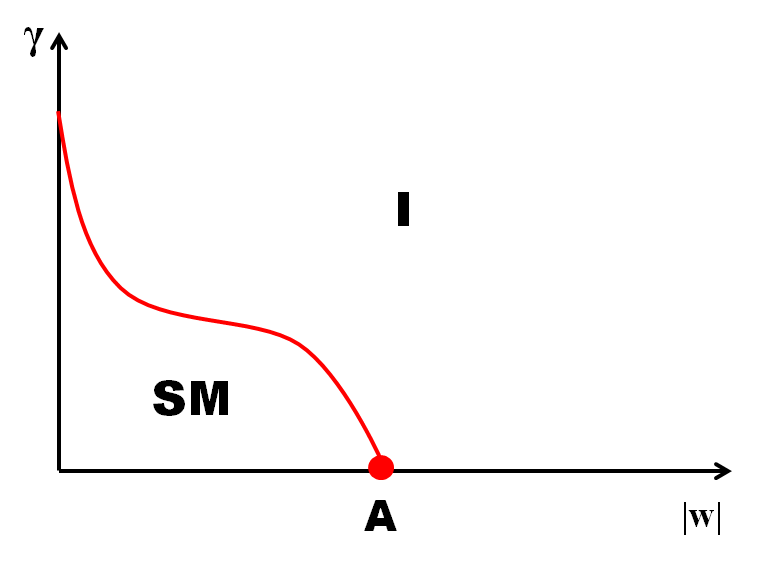}
 \caption{The schematic phase diagram of a non-interacting tilted DSM in the presence of $y$-RVP.
 $\gamma$ and $w$ are the (dimensionless) disorder strength and the tilting parameter, respectively.
 Point A located at $|w|=1$ and $\gamma=0$ is the Lifshitz transition point, separating the type-I and
 type-II DSMs in the absence of disorder.}
 \label{wldff21}
\end{center}
\end{figure}

Now we consider the type-II DSMs. To the one-loop order, we find that $w^{\prime}=w$. It we choose $z$ to
be
\begin{equation}
 z=1+\frac{\Delta u_2^2w(w-\lambda)}{2\pi^2v_1v_2|w|(w^2-1)}w(w-\lambda) \ , \label{wf2rgsol24}
\end{equation}
then both $v_1$ and $v_2$ are RG invariants, and
\begin{eqnarray*}
 \lambda^{\prime} &=& \lambda+\frac{\Delta u_2^2(1-w\lambda)(w-\lambda)}{2\pi^2v_1v_2|w|(w^2-1)}l
 +O(\lambda^2) \ , \\
 u_2^{\prime} &=& u_2+\frac{\Delta u_2^3}{2\pi^2v_1v_2|w|}l+O(\lambda^2) \ .
\end{eqnarray*}
As a result, $w$ and $v_{1,2}$ are marginal to the one-loop order. On the other hand, the one-loop RG
equations for $\lambda$ and $u_2$ are
\begin{eqnarray}
 \frac{d\lambda_l}{dl} &=& \frac{\gamma_{2l}(1-w\lambda)(w-\lambda)}{2(w^2-1)} \ ,
 \label{wf2rge24} \\
 \frac{d\gamma_{2l}}{dl} &=& \gamma_{2l}^2 \ , \label{wf2rge25}
\end{eqnarray}
where $\gamma_{2l}=\frac{\Delta}{|w|\pi^2}(u_{2l}/v)^2$ and for simplicity, we have set $v_1=v=v_2$.
From Eq. (\ref{wf2rge25}), we see that the $u_2$ term is a relevant perturbation. That is, the pure
type-II DSM is unstable in the presence of weak $y$-RVP, and the ground state is supposed to be an
insulator. This is in contrast with type-I DSMs in the presence of weak $y$-RVP. Consequently, we
expect the occurrence of a QPT upon varying the value of $|w|$ for a given disorder strength. A
schematic phase diagram in the presence of $y$-RVP is shown in Fig. \ref{wldff21}. The phase boundary
between the SM and insulator is obtained from Fig. \ref{wldff24}. According to the mean-field theory,
the SM-insulator transition is continuous.

\section{The RM}
\label{rm}
\subsection{Type-I DSMs}

Finally, we consider the RM. To the one-loop order, we find that $w^{\prime}=w$. If we choose $z$ to be
\begin{equation}
 z=1+\frac{\Delta u_3^2}{2\pi v_1v_2(1-w^2)^{3/2}}(1-w\lambda) \ , \label{wf2rg16}
\end{equation}
then both $v_1$ and $v_2$ are RG invariants, and
\begin{eqnarray*}
 \lambda^{\prime} &=& \lambda+\frac{\Delta u_3^2}{2\pi v^2(1-w^2)^{3/2}}(1-w\lambda)(w-\lambda)l \\
 & & +O(l^2) \ , \\
 u_3^{\prime} &=& u_3-\frac{\Delta u_3^3}{2\pi v^2\sqrt{1-w^2}}l+O(l^2) \ .
\end{eqnarray*}
In the last two equations, we have set $v_1=v_2=v$ for simplicity. Hence, the one-loop RG equations for
$\lambda$ and $u_3$ are given by
\begin{equation}
 \frac{d\lambda_l}{dl}=\frac{\Delta u_3^2}{2\pi v^2(1-w^2)^{3/2}}(1-w\lambda_l)(w-\lambda_l) \ ,
 \label{wf2rge16}
\end{equation}
and
\begin{equation}
 \frac{du_{3l}}{dl}=-\frac{\Delta u_{3l}^3}{2\pi v^2\sqrt{1-w^2}} \ , \label{wf2rge12}
\end{equation}
respectively. Equation (\ref{wf2rge12}) has only one fixed point $u_3=0$, with $z=1$. Since the right
hand side in Eq. (\ref{wf2rge12}) is negative, this fixed point is IR stable. In other words, the RM
term is marginally irrelevant at weak disorder. Consequently, the type-I DSM is stable against the weak
RM disorder.

To determine the fate of $\lambda_l$, we have to solve Eqs. (\ref{wf2rge16}) and (\ref{wf2rge12}). By
introducing the dimensionless coupling $\gamma_{3l}=\frac{\Delta}{\pi\sqrt{1-w^2}}(u_{3l}/v)^2$, the
solution with the bare value $\lambda=0$ is
\begin{equation}
 \frac{\lambda_l-w}{\lambda_l-1/w}=\frac{w^2}{\sqrt{1+\gamma_3l}} \ . \label{wf2rgsol15}
\end{equation}
From Eq. (\ref{wf2rgsol15}), we find that $\lambda_*=\lambda_{+\infty}=w$. Using this value of
$\lambda_l$, the dispersion relation of quasiparticles near the Dirac point is given by
\begin{equation}
E_{\pm}(\bm{p})=\pm\sqrt{v_1^2p_1^2+\frac{v_2^2}{1-w^2}p_2^2} \ . \label{wf2rgsol14}
\end{equation}
We see that the quasiparticles can be described by the Dirac fermions with an untilted and anisotropic
Dirac cone. This is consistent with Ref. \onlinecite{ZKYang}.

To sum up, the ground state at $|w|<1$ in the presence of a weak RM disorder is an untilted DSM with an
anisotropic Dirac cone. Since the fermion-disorder coupling is marginally irrelevant, the physical
quantities may acquire logarithmic temperature dependence at low temperatures.

\subsection{Type-II DSMs}

\begin{figure}
\begin{center}
 \includegraphics[width=0.9\columnwidth]{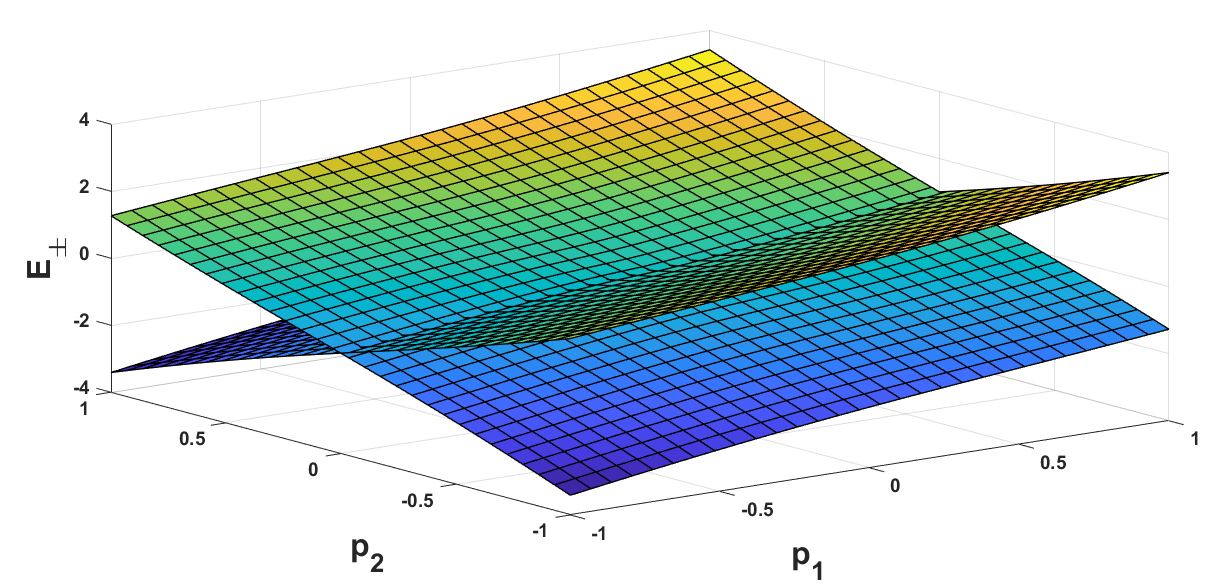}
 \caption{The dispersion relations $E_{\pm}(\bm{p})$ of quasiparticles around a Dirac node with the
 valley index $\xi=1$ in a type-II DSM in the presence of RM, with $w=1.1$ and $z=1.2$. $E_{\pm}(\bm{p})$
 are measured in units of $|w|vk_0$ and the momentum $\bm{p}$ is measured in units of $k_0$. For
 simplicity, we have set $v_1=v=v_2$.}
 \label{wldff13}
\end{center}
\end{figure}

Now we consider type-II DSMs. To the one-loop order, we find that $w^{\prime}=w$. If we choose $z$ to be
\begin{equation}
 z=1+\frac{\Delta u_3^2w(w-\lambda)}{2\pi^2v_1v_2|w|(w^2-1)} \ , \label{wf2rgsol25}
\end{equation}
then we get $v_{1,2}^{\prime}=v_{1,2}$, $u_3^{\prime}=u_3$, and
\begin{eqnarray*}
 \lambda^{\prime}=\lambda+\frac{\Delta u_3^2(1-w\lambda)(w-\lambda)}{2\pi^2v_1v_2|w|(w^2-1)}l
 +O(\lambda^2) \ .
\end{eqnarray*}
As a result, $w$, $v_{1,2}$, and $u_3$ are all marginal to the one-loop order. On the other hand, the
one-loop RG equation for $\lambda$ is
\begin{equation}
 \frac{d\lambda_l}{dl}=\frac{\Delta u_3^2(1-w\lambda)(w-\lambda)}{2\pi^2v_1v_2|w|(w^2-1)} \ .
 \label{wf2rge26}
\end{equation}

The solution of Eq. (\ref{wf2rge26}) with the initial value $\lambda=0$ is
\begin{equation}
 \frac{\lambda_l-1/w}{\lambda_l-w}=\frac{1}{w^2}
 \exp{\! \left(-\frac{\Delta u_3^2l}{2\pi^2v_1v_2|w|}\right)} . \label{wf2rgsol26}
\end{equation}
Hence, we get $\lambda_*=\lambda_{+\infty}=1/w$. Inserting this value of $\lambda_l$ into Eq.
(\ref{wf2rgsol25}), we obtain the dynamical exponent $z=1+\eta$, where
\begin{equation}
 \eta=\frac{\Delta u_3^2}{2\pi^2v_1v_2|w|} \ . \label{wf2rgsol27}
\end{equation}
For $\lambda=\lambda_*$, the dispersion relation of quasiparticles is
\begin{equation}
 E_{\pm}(\bm{p})=w \! \left(\frac{p}{k_0}\right)^{\! \eta} \! \! \left(\xi v_1p_1\pm\frac{v_2}
 {\sqrt{w^2-1}}p_2\right) , \label{wf2rgsol28}
\end{equation}
and $k_0\sim 1/a_0$. A typical form of $E_{\pm}(\bm{p})$ is plotted in Fig. \ref{wldff13}. We see that
the quasiparticles at low energies are not described by the Dirac fermions anymore. However, it is still
a FL with an open Fermi surface given by
\begin{equation}
 p_2=\pm\xi\sqrt{w^2-1}(v_1/v_2)p_1 \ , \label{wf2df1}
\end{equation}
which consists of two straight lines for each valley.

To determine the fate of this FL in the presence of a weak marginal fermion-disorder coupling, we
determine the physical properties in terms of the perturbation theory. This is valid when the disorder
strength is weak since the fermion-disorder coupling is marginal. In particular, we calculate the
one-loop fermion self-energy:
\begin{widetext}
\begin{eqnarray*}
 \Sigma_{\xi\sigma}(ip_0) &=& -\Delta u_3^2 \! \int \! \! \frac{d^2p}{(2\pi)^2}
 \frac{ip_0-\xi wv(p/k_0)^{\eta}p_1}
 {[p_0+i\xi wv(p/k_0)^{\eta}p_1]^2+[-ip_0/w+\xi v(p/k_0)^{\eta}p_1]^2+v^2(p/k_0)^{2\eta}p^2_2} \\
 & & -\Delta u_3^2\sigma_1 \! \int \! \! \frac{d^2p}{(2\pi)^2}\frac{ip_0/w-\xi v(p/k_0)^{\eta}p_1}
 {[p_0+i\xi wv(p/k_0)^{\eta}p_1]^2+[-ip_0/w+\xi v(p/k_0)^{\eta}p_1]^2+v^2(p/k_0)^{2\eta}p^2_2} \ .
\end{eqnarray*}
By analytic continuation $ip_0\rightarrow p_0+i0^+$, the retarded slef-energy $\Sigma^r_{\xi\sigma}(p_0)$
is given by
\begin{eqnarray*}
 \Sigma^r_{\xi\sigma}(p_0) &=& \frac{\Delta u_3^2k_0}{8\pi^2v(1-1/w^2)} \! \left(1+\frac{\sigma_1}{w}
 \right) \! \! \int \! \! d^2x\frac{1}
 {\tilde{p}_0-\xi wr^{\eta}x_1-\sqrt{\frac{w^2}{w^2-1}}r^{\eta}x_2+i0^+} \\
 & & +\frac{\Delta u_3^2k_0}{8\pi^2v(1-1/w^2)} \! \left(1+\frac{\sigma_1}{w}\right) \! \! \int \! \!
 d^2x\frac{1}{\tilde{p}_0-\xi wr^{\eta}x_1+\sqrt{\frac{w^2}{w^2-1}}r^{\eta}x_2+i0^+} \ ,
\end{eqnarray*}
where $x_i=p_i/k_0$ with $i=1,2$, $r=\sqrt{x_1^2+x_2^2}$, and $\tilde{p}_0=p_/(vk_0)$. For simplicity,
we have set $v_1=v=v_2$. As a result, its imaginary part is of the form
\begin{eqnarray*}
 \mbox{Im}\Sigma^r_{\xi\sigma}(p_0) &=& -\frac{\Delta u_3^2k_0}{8\pi v(1-1/w^2)} \! \left(
 1+\frac{\sigma_1}{w}\right) \! \! \int \! \! d^2x\delta \! \left(\tilde{p}_0-\xi wr^{\eta}x_1
 -\sqrt{\frac{w^2}{w^2-1}}r^{\eta}x_2\right) \\
 & & -\frac{\Delta u_3^2k_0}{8\pi v(1-1/w^2)} \! \left(1+\frac{\sigma_1}{w}\right) \! \! \int \! \!
 d^2x\delta \! \left(\tilde{p}_0-\xi wr^{\eta}x_1+\sqrt{\frac{w^2}{w^2-1}}r^{\eta}x_2\right) .
\end{eqnarray*}
Setting $\tilde{p}_0=0$, we find that
\begin{eqnarray*}
 \mbox{Im}\Sigma^r_{\xi\sigma}(0) &=& -\frac{\Delta u_3^2k_0|w|}{8\pi v(w^2-1)} \! \left(
 1+\frac{\sigma_1}{w}\right) \! \! \int \! \! d^2x\frac{1}{r^{\eta}}\delta \! \left(x_1+\frac{\eta_w\xi}
 {\sqrt{w^2-1}}x_2\right) \\
 & & -\frac{\Delta u_3^2k_0|w|}{8\pi v(w^2-1)} \! \left(1+\frac{\sigma_1}{w}\right) \! \! \int \! \!
 d^2x\frac{1}{r^{\eta}}\delta \! \left(x_1-\frac{\eta_w\xi}{\sqrt{w^2-1}}x_2\right) \\
 &=& -\frac{\Delta u_3^2k_0|w|^{1-\eta}}{2\pi v(w^2-1)^{1-\eta/2}} \! \left(1+\frac{\sigma_1}{w}\right)
 \! \! \int^{+\infty}_0 \! \frac{dx_2}{x_2^{\eta}} \ .
\end{eqnarray*}
\end{widetext}
Since $\eta<1$ for the weak disorder strength, the integral is UV divergent. We have to cut it off at a
scale $D/(vk_0)$ where $D=O(vk_0)$ is the band width. Without loss of generality, we choose $D=vk_0$.
(A different choice of the ratio $D/(vk_0)$ corresponds to the redefinition of the bare value $u_3^2$.)
Thus, we have
\begin{eqnarray*}
 \mbox{Im}\Sigma^r_{\xi\sigma}(0) &=& -\frac{\Delta u_3^2k_0|w|^{1-\eta}}{2\pi v(w^2-1)^{1-\eta/2}} \!
 \left(1+\frac{\sigma_1}{w}\right) \! \! \int^1_0 \! \frac{dx_2}{x_2^{\eta}} \\
 &=& -\frac{\Delta u_3^2k_0|w|^{1-\eta}}{2\pi v(1-\eta)(w^2-1)^{1-\eta/2}} \! \left(1+\frac{\sigma_1}{w}
 \right) .
\end{eqnarray*}

This result implies that to the one-loop order, the single-particle Green function for quasiparticles at
low frequencies and small momenta is of the form
\begin{eqnarray}
 G_{\xi}^{-1}(P) &=& \! \left(1+\frac{\sigma_1}{w}\right) \! \! \left[-ip_0+\xi v \! \left(\frac{p}{k_0}
 \right)^{\! \eta} \! p_1-\frac{\mbox{sgn}(p_0)}{2\tau}\right] \nonumber \\
 & & +v \! \left(\frac{p}{k_0}\right)^{\! \eta} \! p_2\sigma_2 \ , \label{wf2rm1}
\end{eqnarray}
in the imaginary-time formulation, where the mean free time $\tau$ is given by
\begin{equation}
 \frac{1}{\tau}=\frac{\Delta u_3^2k_0|w|^{1-\eta}}{\pi v(1-\eta)(w^2-1)^{1-\eta/2}} \ . \label{wf2rm11}
\end{equation}
Since the quasiparticles acquire a nonzero mean free time, the system is a DM at weak disorder strength.
According to the scaling theory of localization\cite{Abrahams}, a DM phase in $d=2$ is unstable in the
presence of weak disorder and turns into an insulator. Alternatively, we can investigate the role of the
marginal fermion-disorder coupling by a replica mean-field theory, similar to what we have done for
type-I DSMs in the presence of $y$-RVP. The above perturbative calculation suggests that the mean-field
equation always has a non-zero solution such that the quasiparticles acquire a nonvanishing mean free
time. The fluctuations around this broken-symmetry solution are described by a generalized nonlinear
$\sigma$ model. In two dimensions, the nonlinear $\sigma$ model has only one phase -- the disordered
phase, corresponding to the insulator within the present context. In any case, we reach the conclusion
that the ground state is insulating for $|w|>1$.

\begin{figure}
\begin{center}
 \includegraphics[width=0.9\columnwidth]{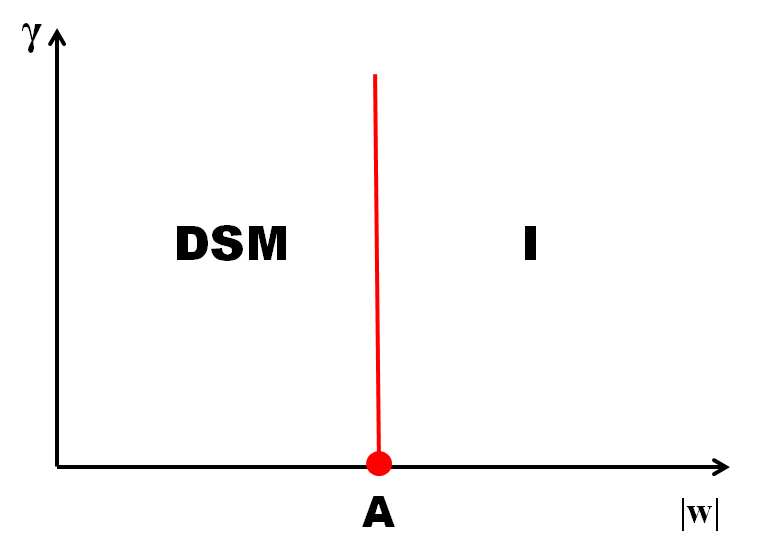}
 \caption{The schematic phase diagram of a non-interacting tilted DSM in the presence of weak RM.
 $\gamma$ and $w$ are the (dimensionless) disorder strength and the tilting parameter, respectively.
 Point A located at $|w|=1$ and $\gamma=0$ is the Lifshitz transition point, separating the type-I and
 type-II DSMs in the absence of disorder. The phase boundary between the DSM and insulating phase is
 schematic. When the disorder strength is further increased, the DSM will turn into an insulator.}
 \label{wldff22}
\end{center}
\end{figure}

Since the system with $|w|<1$ is an untilted DSM at weak disorder strength and insulating when $|w|>1$,
we conclude that there is a QPT from $|w|<1$ to $|w|>1$ for a given disorder strength in the presence of
weak RM. The resulting schematic phase diagram is shown in Fig. \ref{wldff22}. In fact, as we approach
the phase boundary between the DSM and the insulating phase from the side of the DSM, the component of
the velocity perpendicular to the tilting direction (the $y$-direction in the present setup) becomes
singular at the phase boundary. On the other hand, if we approach the phase boundary from the side of
the insulator, we find that $1/\tau\rightarrow +\infty$ at the phase boundary. All these imply that the
quantum fluctuations are strong around the line $|w|=1$ and the starting point we have adopted, i.e.,
starting from either $|w|<1$ or $|w|>1$ may not be appropriate. As a result, Fig. \ref{wldff22} is just
schematic, and the exact location of the phase boundary may not be a straight line. Moreover, other
phases may exist close to the $|w|=1$ line.

\section{Conclusions and discussions}

We study the effects of various types of quenched disorder on the non-interacting tilted Dirac fermions
in two dimensions with the help of the perturbative RG. Since the RG transformations must scale to the
Fermi surface, we parametrize the low-energy degrees of freedom according to their energies so that we
can integrate out the modes with large energies properly. When the Fermi surface is point-like, the
results are consistent with those by integrating out the modes with large momenta. On the other hand,
the answers may be different when the Fermi surface is extended. Although we focus on the $2$D tilted
DSMs, it is straightforward ro extend our method to the tilted WSMs in three dimensions.

The relevancy of various fermion-disorder couplings under the RG transformations in both types of DSMs
to the one-loop order is summarized in Table \ref{T1}. Whenever the fermion-disorder coupling is
relevant, we extrapolate our one-loop RG equations to the strong disorder regime and claim that the
resulting phase is an insulator. When the fermion-disorder coupling is marginal, we examine its role by
using either the mean-field approximation of a replica field theory or the first-order Born approximation.
If the fermion-disorder coupling is irrelevant, then this phase is stable against the presence of weak
disorder.

When $w\neq 0$, the RSP and the $x$-RVP will generate each other under the RG transformations even if
one of the bare value is zero. Hence, we must consider them together when calculating the RG equations.
In the presence of the RSP or the $x$-RVP, we find that both types of DSMs become insulators even at weak
disorder strength because the corresponding fermion-disorder coupling flows to strong disorder regime.

\begin{table}
\begin{center}
 \begin{tabular}{|c|c|c|} \hline
 disorder & type-I & type-II \\ \hline
 RSP & relevant & relevant \\ \hline
 $x$-RVP & relevant & relevant \\ \hline
 $y$-RVP & marginal & relevant \\ \hline
 RM & irrelevant & marginal \\ \hline
 \end{tabular}
 \caption{The relevancy of various fermion-disorder couplings under the RG transformations in both types
 of DSMs to the one-loop order.}
 \label{T1}
\end{center}
\end{table}

For the $y$-RVP, we find that the system at $|w|<1$ is a SM characterized by a non-universal dynamical
exponent $z>1$. This SM is fragile since it becomes an insulator at a moderate strength of disorder.
Especially, the critical disorder strength vanishes as $|w|\rightarrow 1$. On the other hand, the system
is insulating at $|w|>1$. Thus, we expect that there is a SM-insulator transition upon varying $|w|$ for
a given disorder strength, which is continuous according to our replica mean-field theory. The
calculations of the critical exponents associated with this transition are beyond the scope of the
present work.

For the weak RM, the system at $|w|<1$ is an untilted DSM with an anisotropic Dirac cone. On the other
hand, it is an insulator when $|w|>1$. Thus, we expect that there is a DSM-insulator transition upon
varying $|w|$ for a given disorder strength. The determination of the nature of this transition is beyond
the scope of the present work. Moreover, on account of the strong fluctuations close to the $|w|=1$ line,
our approach which starts from either side may fail, and there can exist other phases near the $|w|=1$
line.

For type-I DSMs, the effects of the quenched disorder have been studied with a different type of RG
scheme\cite{ZKYang}. For the $y$-RVP and RM, the phases we find at the weak disorder are identical to the
ones in Ref. \onlinecite{ZKYang}. For the former, we indicate that the SM may be unstable upon increasing
the disorder strength, which has not been examined in Ref. \onlinecite{ZKYang}. We further determine the
critical disorder strength beyond which the SM becomes unstable toward an insulator. The main difference
between our work and Ref. \onlinecite{ZKYang} lies on the nature of the ground state of type-I DSMs in
the presence of RSP or $x$-RVP. According to Ref. \onlinecite{ZKYang}, the ground state is a DM with a
bulk Fermi arc. This DM cannot be stable since the fermion-disorder coupling flows to the strong disorder
regime at low energies. One possibility in the strong disorder regime is an insulating phase due to the
random potential scattering.

Further studies, maybe numerics, are warranted to justify the phase diagrams we have obtained in this
work. In particular, the nature of the DSM-insulator transition and the phases close to the transition
line in the presence of a weak RM are open questions. Since electrons carry electric charges, the
long-range Coulomb interaction is always present. It is interesting to investigate how the
electron-electron interactions affect the phase diagrams. For type-I DSMs, this question has been
studied in Ref. \onlinecite{ZKYang}. For type-II DSMs, however, this question remains unanswered.

\acknowledgments

The works of Y.-W. Lee is supported by the Ministry of Science and Technology, Taiwan,
under the grant number MOST 107-2112-M-029-002.


\appendix
\section{Derivation of the one-loop RG equations}
\label{rge}

\begin{figure}
\begin{center}
 \includegraphics[width=0.9\columnwidth]{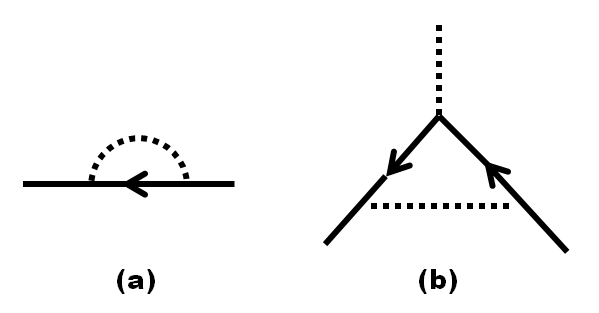}
 \caption{The one-loop correction to the self-energy of Dirac fermions (a) and the fermion-disorder
 coupling (b). The solid and the dashed lines correspond to the fermion propagator and the disorder
 potential, respectively.}
 \label{wldff1}
\end{center}
\end{figure}

Here we present the details of the derivation of the one-loop RG equations. To properly integrate out
the modes with large energies, we have parametrized the low-energy degrees of freedom according to their
energies, as shown in Eqs (\ref{wf2e1}) and (\ref{wf2e11}) for type-I and type-II DSMs, respectively. In
terms of them, we write the involved momentum integrals as
\begin{eqnarray}
 \! \int \! d^2\tilde{p} \! \! &=& \! \! \frac{1}{2} \! \int^{\Lambda}_0 \! \frac{EdE}{(1-w^2)^{3/2}} \!
 \int^{2\pi}_0 \! \!d\theta(1-\xi w\cos{\theta}) \nonumber \\
 \! \! & & \! \! +\frac{1}{2} \! \int_{-\Lambda}^0 \! \frac{|E|dE}{(1-w^2)^{3/2}} \! \int^{2\pi}_0 \! \!
 d\theta(1+\xi w\cos{\theta}) , ~~~~~~~~\label{wf2e12}
\end{eqnarray}
for type-I Dirac fermions, and
\begin{eqnarray}
 \! \int \! d^2\tilde{p} &=& \frac{|w|}{2} \! \int^{\Lambda}_0 \! \frac{EdE}{(w^2-1)^{3/2}} \! \left[\!
 \int^{+\infty}_{-\infty} \! d\theta(\cosh{\theta}+\xi/w)\right. \nonumber \\
 & & +\! \left. \! \int^{+\infty}_{-\infty} \! d\theta(\cosh{\theta}-\xi/w)\right] \nonumber \\
 & & +\frac{|w|}{2} \! \int_{-\Lambda}^0 \! \frac{|E|dE}{(w^2-1)^{3/2}} \! \left[\!
 \int^{+\infty}_{-\infty} \! d\theta(\cosh{\theta}-\xi/w)\right. \nonumber \\
 & & +\! \left. \! \int^{+\infty}_{-\infty} \! d\theta(\cosh{\theta}+\xi/w)\right] , \label{wf2e13}
\end{eqnarray}
for type-II Dirac fermions, where $\tilde{p}_i=v_ip_i$ with $i=1,2$ and $\Lambda$ is the UV cutoff in
energies. In Eq. (\ref{wf2e13}), the first and the second $\theta$ integrals for given $E$ correspond to
the integrations over the right and the left branches of the hyperbola, respectively. In fact, it
suffices to consider the integrals over $E>0$ or $E<0$ since the involved two bands have been taken into
account by the Pauli matrices. However, this regularization breaks the PH symmetry of $H_0$ at $\mu=0$.
Hence, we define the momentum integrals by Eqs. (\ref{wf2e12}) or (\ref{wf2e13}). This accounts for the
prefactor $1/2$.

There are only two diagrams which contribute to the one-loop RG equations, i.e., the self-energy of Dirac
fermions and the vertex correction to the fermion-disorder coupling, as illustrated in Fig. \ref{wldff1}.
We discuss them separately in the following.

The one-loop self-energy of Dirac fermions is given by
\begin{widetext}
\begin{eqnarray*}
 & & \Sigma_{\xi\sigma}(K)=-\frac{1}{2v_1v_2}\Delta\cdot 2\cdot \! \int_{\mathcal{D}} \!
     \frac{d^2\tilde{p}}{(2\pi)^2}\Gamma G_{\xi}^{(0)}(ik_0,\bm{p})\Gamma \\
 & & =-\frac{\Delta}{v_1v_2} \! \int_{\mathcal{D}} \! \frac{d^2\tilde{p}}{(2\pi)^2}
	 \frac{(ik_0-\xi w\tilde{p}_1)\Gamma^2}
	 {(k_0+i\xi w\tilde{p}_1)^2+(-i\lambda k_0+\xi\tilde{p}_1)^2+\tilde{p}^2_2}
	 -\frac{\Delta}{v_1v_2} \! \int_{\mathcal{D}} \! \frac{d^2\tilde{p}}{(2\pi)^2}
	 \frac{\Gamma[(-i\lambda k_0+\xi\tilde{p}_1)\sigma_1+\tilde{p}_2\sigma_2]\Gamma}
	 {(k_0+i\xi w\tilde{p}_1)^2+(-i\lambda k_0+\xi\tilde{p}_1)^2+\tilde{p}^2_2} \\
 & & =-\frac{\Delta}{v_1v_2} \! \int_{\mathcal{D}} \! \frac{d^2\tilde{p}}{(2\pi)^2}
	 \frac{(ik_0-\xi w\tilde{p}_1)\Gamma^2}
	 {(k_0+i\xi w\tilde{p}_1)^2+(-i\lambda k_0+\xi\tilde{p}_1)^2+\tilde{p}^2_2}
	 -\frac{\Delta}{v_1v_2} \! \int_{\mathcal{D}} \! \frac{d^2\tilde{p}}{(2\pi)^2}
	 \frac{(-i\lambda k_0+\xi\tilde{p}_1)\Gamma\sigma_1\Gamma}
	 {(k_0+i\xi w\tilde{p}_1)^2+(-i\lambda k_0+\xi\tilde{p}_1)^2+\tilde{p}^2_2} \ .
\end{eqnarray*}
where $k_0$ and $\bm{k}$ are, respectively, the external frequency and the external momentum,
$\mathcal{D}$ denotes the energy shell in the range $\Lambda/s<|E|<\Lambda$, and
\begin{eqnarray*}
 G_{\xi}^{(0)}(ip_0,\bm{p})=\frac{1}
 {-ip_0+\xi w\tilde{p}_1+(-i\lambda p_0+\xi\tilde{p}_1)\sigma_1+\tilde{p}_2\sigma_2} \ .
\end{eqnarray*}
We will take $l\rightarrow 0$ at the end of calculations. The last equality follows from the facts that
$\mathcal{D}$ is symmetric under the reflection $\tilde{p}_2\rightarrow -\tilde{p}_2$. We see that
$\Sigma_{\xi\sigma}(K)$ depends only on $k_0$ to the one-loop order, and we will denote it by
$\Sigma_{\xi\sigma}(ik_0)$. By performing the derivative expansion, we have for the RSP or $x$-RVP
\begin{eqnarray*}
 \Sigma_{\xi\sigma}(ik_0) \! \! &=& \! \! -\frac{\Delta}{v_1v_2}ik_0I_1(u_0^2+u_1^2-2\lambda u_0u_1-\sigma_1
 [\lambda(u_0^2+u_1^2)-2u_0u_1]) \\
 & & +\frac{\Delta}{v_1v_2}\xi I_2[w(u_0^2+u_1^2)-2u_0u_1-\sigma_1(u_0^2+u_1^2-2wu_0u_1)] \\
 & & -\frac{\Delta}{v_1v_2}2(w-\lambda)ik_0I_3[w(u_0^2+u_1^2)-2u_0u_1-\sigma_1(u_0^2+u_1^2-2wu_0u_1)]
 +O(k_0^2) \ ,
\end{eqnarray*}
and for the $y$-RVP or RM,
\begin{eqnarray*}
 \Sigma_{\xi\sigma}(ik_0)=-\frac{\Delta u^2_{2/3}}{v_1v_2}[ik_0(1+\lambda\sigma_1)I_1-\xi(w+\sigma_1)I_2
 +2(w-\lambda)ik_0(w+\sigma_1)I_3]+O(k_0^2) \ ,
\end{eqnarray*}
\end{widetext}
where
\begin{eqnarray*}
 I_1 &=& \! \int_{\mathcal{D}} \! \frac{d^2\tilde{p}}{(2\pi)^2}\frac{1}{(1-w^2)\tilde{p}^2_1+\tilde{p}^2_2}
 \ , \\
 I_2 &=& \! \int_{\mathcal{D}} \! \frac{d^2\tilde{p}}{(2\pi)^2}\frac{\tilde{p}_1}
 {(1-w^2)\tilde{p}^2_1+\tilde{p}^2_2} \ , \\
 I_3 &=& \! \int_{\mathcal{D}} \! \frac{d^2\tilde{p}}{(2\pi)^2}\frac{\tilde{p}_1^2}
 {[(1-w^2)\tilde{p}^2_1+\tilde{p}^2_2]^2} \ .
\end{eqnarray*}

The one-loop correction $\delta S_{\Gamma}$ to the fermion-disorder coupling is given by
\begin{eqnarray*}
 & & \delta S_{\Gamma} \\
 & & =-\frac{\Delta}{v_1v_2} \! \int_{\mathcal{D}} \! \frac{d^2\tilde{p}}{(2\pi)^2}\Gamma G_{\xi}^{(0)}
     (ik_0,\bm{p}+\bm{k})\Gamma G_{\xi}^{(0)}(ik_0,\bm{p})\Gamma \\
 & & =-\frac{\Delta}{v_1v_2} \! \int_{\mathcal{D}} \! \frac{d^2\tilde{p}}{(2\pi)^2}\Gamma G_{\xi}^{(0)}
     (0,\bm{p})\Gamma G_{\xi}^{(0)}(0,\bm{p})\Gamma+\cdots \ ,
\end{eqnarray*}
where $\cdots$ denotes the higher-order terms in powers of $\bm{k}$ and $k_0$, which will be ignored
hereafter. For the RSP or $x$-RVP, we have
\begin{eqnarray*}
 \delta S_{\Gamma} &=& -\frac{\Delta u_0^3}{v_1v_2}(I_1+2w^2I_3-2wI_3\sigma_1) \\
 & & +\frac{\Delta u_1^3}{v_1v_2}[2wI_3+(I_1-2I_3)\sigma_1] \\
 & & +\frac{\Delta u^2_0u_1}{v_1v_2}[2wI_3+(I_1-2I_3)\sigma_1] \\
 & & -\frac{2\Delta u_0^2u_1}{v_1v_2}[-2wI_3+(I_1+2w^2I_3)\sigma_1] \\
 & & -\frac{\Delta u^2_1u_0}{v_1v_2}(I_1+2w^2I_3-2wI_3\sigma_1) \\
 & & +\frac{2\Delta u^2_1u_0}{v_1v_2}(I_1-2I_3+2wI_3\sigma_1) \ .
\end{eqnarray*}
For the $y$-RVP and the RM, we find that
\begin{eqnarray*}
 \delta S_{\Gamma}=\frac{\Delta u_2^3}{v_1v_2}\sigma_2(I_1-2I_4) \ ,
\end{eqnarray*}
and
\begin{eqnarray*}
 \delta S_{\Gamma}=\frac{\Delta u_3^3}{v_1v_2}\sigma_3I_1 \ ,
\end{eqnarray*}
respectively, where
\begin{eqnarray*}
 I_4=\! \int_{\mathcal{D}} \! \frac{d^2\tilde{p}}{(2\pi)^2}\frac{\tilde{p}_2^2}
 {[(1-w^2)\tilde{p}^2_1+\tilde{p}^2_2]^2} \ .
\end{eqnarray*}

The rest of the task is to calculate the four integrals $I_1,\cdots,I_4$. The answers depend on the type
of Dirac fermions. We will calculate them separately in the following.

\subsection{Type-I DSMs}

We first consider type-I DSMs. In this case, we have
\begin{eqnarray*}
 I_1 &=& \frac{l}{8\pi^2\sqrt{1-w^2}}[F_1(\xi,w)+F_1(-\xi,w)] \ , \\
 I_2 &=& \frac{l\Lambda}{8\pi^2(1-w^2)^{3/2}}[F_2(\xi,w)+F_2(-\xi,w)]+O(l^2) \ , \\
 I_3 &=& \frac{l}{8\pi^2(1-w^2)^{3/2}}[F_3(\xi,w)+F_3(-\xi,w)] \ , \\
 I_4 &=& \frac{l}{8\pi^2\sqrt{1-w^2}}[F_4(\xi,w)+F_4(-\xi,w)] \ ,
\end{eqnarray*}
where
\begin{eqnarray*}
 F_1(\xi,w) &=& \! \int^{2\pi}_0 \! d\theta\frac{1-\xi w\cos{\theta}}{w^2-2\xi w\cos{\theta}+1} \\
 &=& 2\pi \ , \\
 F_2(\xi,w) &=& \! \int^{2\pi}_0 \! d\theta\frac{(-\xi w+\cos{\theta})(1-\xi w\cos{\theta})}
 {w^2-2\xi w\cos{\theta}+1} \\
 &=& -\pi\xi w \ , \\
 F_3(\xi,w) &=& \! \int^{2\pi}_0 \! d\theta\frac{(-\xi w+\cos{\theta})^2(1-\xi w\cos{\theta})}
 {(w^2-2\xi w\cos{\theta}+1)^2} \\
 &=& \pi \ , \\
 F_4(\xi,w) &=& \! \int^{2\pi}_0 \! d\theta\frac{\sin^2{\theta}(1-\xi w\cos{\theta})}
 {(w^2-2\xi w\cos{\theta}+1)^2} \\
 &=& \pi \ .
\end{eqnarray*}
Consequently, we get
\begin{eqnarray*}
 I_1 &=& \frac{l}{2\pi\sqrt{1-w^2}} \ , \\
 I_2 &=& 0 \ , \\
 I_3 &=& \frac{l}{4\pi(1-w^2)^{3/2}} \ , \\
 I_4 &=& \frac{l}{4\pi\sqrt{1-w^2}} \ .
\end{eqnarray*}

For the RSP or $x$-RVP, we find that
$\Sigma_{\xi\sigma}(i\omega)=(-i\omega)(\Sigma_0\sigma_0+\Sigma_1\sigma_1)$, where
\begin{eqnarray*}
 \Sigma_0 &=& \frac{\Delta(1-w\lambda)}{2\pi v_1v_2(1-w^2)^{3/2}}(u_0^2+u_1^2-2wu_0u_1)l \ , \\
 \Sigma_1 &=& -\frac{\Delta(1-w\lambda)}{2\pi v_1v_2(1-w^2)^{3/2}}[w(u_0^2+u_1^2)-2u_0u_1]l \ ,
\end{eqnarray*}
and $\delta S_{\Gamma}=V_0\sigma_0+V_1\sigma_1$, where
\begin{eqnarray*}
 V_0 &=& -\frac{\Delta[u_0^3-wu_1^3-3wu_0^2u_1+(1+2w^2)u_1^2u_0]}{2\pi v_1v_2(1-w^2)^{3/2}}l \ , \\
 V_1 &=& \frac{\Delta [wu_0^3-w^2u_1^3+3wu_1^2u_0-(2+w^2)u_0^2u_1]}{2\pi v_1v_2(1-w^2)^{3/2}}l \ .
\end{eqnarray*}
Consequently, the Lagrangian density for the slow modes to the one-loop order is of the form
\begin{eqnarray*}
 \mathcal{L} &=& \! \sum_{\xi,\sigma}\psi^{\dagger}_{\xi\sigma<}[(1+\Sigma_0)+(\lambda+\Sigma_1)\sigma_1]
 \partial_{\tau}\psi_{\xi\sigma<} \\
 & & -\! \sum_{\xi,\sigma}\psi^{\dagger}_{\xi\sigma<}[i\xi v_1(w+\sigma_1)\partial_1+iv_2\sigma_2
 \partial_2]\psi_{\xi\sigma<} \\
 & & -\! \sum_{j=0,1}(u_j-V_j) \! \sum_{\xi,\sigma} \psi_{\xi\sigma<}^{\dagger}\sigma_j\psi_{\xi\sigma<}
 A(\bm{r}) \ .
\end{eqnarray*}

We rescale the variables and fields according to Eq. \ref{wf2rg12} to bring the term $\psi^{\dagger}_{\xi\sigma<}\partial_{\tau}\psi_{\xi\sigma<}$ back to the original form. Then, we have
\begin{equation}
 Z_{\psi}=e^{2l}(1+\Sigma_0) \ , \label{wf2rg13}
\end{equation}
and the Lagrangian density becomes
\begin{eqnarray*}
 \mathcal{L} \! \! &=& \! \! \! \sum_{\xi,\sigma}\psi^{\dagger}_{\xi\sigma}[1+(\lambda+\Sigma_1)
 (1+\Sigma_0)^{-1}\sigma_1]\partial_{\tau}\psi_{\xi\sigma} \\
 \! \! & & \! \! -Z_{\psi}^{-1}e^{(z+1)l} \! \sum_{\xi,\sigma}\psi^{\dagger}_{\xi\sigma}[i\xi v_1
 (w+\sigma_1)\partial_1+iv_2\sigma_2\partial_2]\psi_{\xi\sigma} \\
 \! \! & & \! \! -Z_{\psi}^{-1}e^{(z+1)l} \! \sum_{j=0,1}(u_j-V_j) \! \sum_{\xi,\sigma}
 \psi_{\xi\sigma}^{\dagger}\sigma_j\psi_{\xi\sigma}A(\bm{r}) \ .
\end{eqnarray*}
Therefore, the renormalized parameters are given by
\begin{eqnarray*}
 (wv_1)^{\prime} &=& Z_{\psi}^{-1}e^{(z+1)l}wv_1 \ , \\
 v_{1,2}^{\prime} &=& Z_{\psi}^{-1}e^{(z+1)l}v_{1,2} \ , \\
 \lambda^{\prime} &=& (\lambda+\Sigma_1)(1+\Sigma_0)^{-1} \ , \\
 u_{0,1}^{\prime} &=& Z_{\psi}^{-1}e^{(z+1)l}(u_{0,1}-V_{0,1}) \ ,
\end{eqnarray*}
which give the equations in the main text.

For the $y$-RVP,
\begin{eqnarray*}
 \Sigma_0 &=& \frac{\Delta u_2^2}{2\pi v_1v_2(1-w^2)^{3/2}}(1-w\lambda)l \ , \\
 \Sigma_1 &=& \frac{\Delta u_2^2}{2\pi v_1v_2(1-w^2)^{3/2}}w(1-w\lambda)l \ ,
\end{eqnarray*}
and $\delta S_{\Gamma}=0$. On the other hand, for the RM,
\begin{eqnarray*}
 \Sigma_0 &=& \frac{\Delta u_3^2}{2\pi v_1v_2(1-w^2)^{3/2}}(1-w\lambda)l \ , \\
 \Sigma_1 &=& \frac{\Delta u_3^2}{2\pi v_1v_2(1-w^2)^{3/2}}w(1-w\lambda) l \ ,
\end{eqnarray*}
and $\delta S_{\Gamma}=V_3\sigma_3$, where
\begin{eqnarray*}
 V_3=\frac{\Delta u_3^3}{2\pi v_1v_2\sqrt{1-w^2}}l \ .
\end{eqnarray*}
By rescaling the variables and fields according to Eq. \ref{wf2rg12}, we obtained the one-loop RG
equations in the main text.

\subsection{Type-II DSMs}

Next, we consider type-II DSMs. In this case, we have
\begin{eqnarray*}
 I_1 &=& -\frac{1}{8\pi^2} \! \int^{\Lambda}_{\Lambda/s} \! \frac{dE/E}{\sqrt{w^2-1}}
 [K_1(\xi,w)+K_1(-\xi,w)] \\
 & & -\frac{1}{8\pi^2} \! \int_{-\Lambda}^{-\Lambda/s} \! \frac{dE/|E|}{\sqrt{w^2-1}}
 [K_1(\xi,w)+K_1(-\xi,w)] \ , \\
 I_2 &=& -\frac{1}{8\pi^2} \! \int^{\Lambda}_{\Lambda/s} \! \frac{dE}{(w^2-1)^{3/2}}
 [K_2(\xi,w)-K_2(-\xi,w)] \\
 & & -\frac{1}{8\pi^2} \! \int_{-\Lambda}^{-\Lambda/s} \! \frac{dE}{(w^2-1)^{3/2}}
 [K_2(-\xi,w)-K_2(\xi,w)] \\
 &=& 0 \ , \\
 I_3 &=& \frac{1}{8\pi^2} \! \int^{\Lambda}_{\Lambda/s} \! \frac{dE/E}{(w^2-1)^{3/2}}
 [K_3(\xi,w)+K_3(-\xi,w)] \\
 & & +\frac{1}{8\pi^2} \! \int_{-\Lambda}^{-\Lambda/s} \! \frac{dE/|E|}{(w^2-1)^{3/2}}
 [K_3(\xi,w)+K_3(-\xi,w)] \ , \\
 I_4 &=& \frac{1}{8\pi^2} \! \int^{\Lambda}_{\Lambda/s} \! \frac{dE/E}{\sqrt{w^2-1}}
 [K_4(\xi,w)+K_4(-\xi,w)] \\
 & & +\frac{1}{8\pi^2} \! \int_{-\Lambda}^{-\Lambda/s} \! \frac{dE/|E|}{\sqrt{w^2-1}}
 [K_4(\xi,w)+K_4(-\xi,w)] \ ,
\end{eqnarray*}
where
\begin{eqnarray*}
 K_1(\xi,w) &=& \! \int^{+\infty}_{-\infty} \! d\theta\frac{|w|(\cosh{\theta}+\xi/w)}
 {w^2+2\xi w\cosh{\theta}+1} \ , \\
 K_2(\xi,w) &=& \! \int^{+\infty}_{-\infty} \! d\theta
 \frac{|w|(\xi w+\cosh{\theta})(\cosh{\theta}+\xi/w)}{w^2+2\xi w\cosh{\theta}+1} \ , \\
 K_3(\xi,w) &=& \! \int^{+\infty}_{-\infty} \! d\theta\frac{|w|(\xi w+\cosh{\theta})^2(\cosh{\theta}+\xi/w)}
 {(w^2+2\xi w\cosh{\theta}+1)^2} \ , \\
 K_4(\xi,w) &=& \! \int^{+\infty}_{-\infty} \! d\theta\frac{|w|\sinh^2{\theta}(\cosh{\theta}+\xi/w)}
 {(w^2+2\xi w\cos{\theta}+1)^2} \ .
\end{eqnarray*}
We notice that the integrals in $K_1$, $K_3$, and $K_4$ are UV divergent. By introducing the UV cutoff
in $\theta$, denoted by $\theta_{\Lambda}$, we obtain
\begin{eqnarray*}
 & & K_1(\xi,w)=\eta_w\xi(\theta_{\Lambda}-\ln{|w|}) \ , \\
 & & K_3(\xi,w)\approx\frac{e^{\theta_{\Lambda}}}{4|w|}=K_4(\xi,w) \ .
\end{eqnarray*}
Since $K_1$ is an odd function of $\xi$, we get $I_1=0$.

This UV arises from the linear approximation we have made in the Hamiltonian. In real crystal, the size
of the open Fermi surface in type-II DSMs is restricted by that of the first BZ. Hence, the value of
$\theta_{\Lambda}$ is determined by the size of the first BZ. Suppose that the maximum value of $|p_2|$
is $\pi/a_0$ where $a_0$ is the lattice spacing at the scale $l$. Using the parametrization for
$\tilde{p}_2$, we find that for given $E$
\begin{eqnarray*}
 \frac{|E|}{\sqrt{w^2-1}}\frac{e^{\theta_{\Lambda}}}{2}\approx\frac{v_2\pi}{a_0}\equiv D \ ,
\end{eqnarray*}
leading to
\begin{eqnarray*}
 e^{\theta_{\Lambda}}\approx\frac{2\sqrt{w^2-1}D}{|E|} \ .
\end{eqnarray*}
Consequently, we get
\begin{eqnarray*}
 I_3 &=& \frac{Dl}{4\pi^2\Lambda|w|(w^2-1)} \ , \\
 I_4 &=& \frac{Dl}{4\pi^2\Lambda|w|} \ .
\end{eqnarray*}
In general, $\Lambda=O(D)$ at the energy scale $\Lambda$. Without loss of generality, we set $\Lambda=D$
and we get
\begin{eqnarray*}
 I_3=\frac{l}{4\pi^2|w|(w^2-1)} \ , ~~I_4=\frac{l}{4\pi^2|w|} \ .
\end{eqnarray*}
The choice of the ratio $D/\Lambda$ is arbitrary. But it will not affect the low-energy physics.
Different choices correspond to different bare values of fermion-disorder couplings.

For the RSP or $x$-RVP, we find that
$\Sigma_{\xi\sigma}(i\omega)=(-i\omega)(\Sigma_0\sigma_0+\Sigma_1\sigma_1)$, where
\begin{eqnarray*}
 \Sigma_0 &=& \frac{\Delta(w-\lambda)[w(u_0^2+u_1^2)-2u_0u_1]}{2\pi^2v_1v_2|w|(w^2-1)}l \ , \\
 \Sigma_1 &=& -\frac{\Delta(w-\lambda)(u_0^2+u_1^2-2wu_0u_1)}{2\pi^2v_1v_2|w|(w^2-1)}l \ ,
\end{eqnarray*}
and $\delta S_{\Gamma}=V_0\sigma_0+V_1\sigma_1$, where
\begin{eqnarray*}
 V_0 &=& -\frac{\Delta[w^2u_0^3-wu_1^3-3wu_0^2u_1+(w^2+2)u_1^2u_0]}{2\pi^2v_1v_2|w|(w^2-1)}l \\
 V_1 &=& \frac{\Delta[wu_0^3-u_1^3+3wu_1^2u_0-(1+2w^2)u_0^2u_1]}{2\pi^2v_1v_2|w|(w^2-1)}l \ .
\end{eqnarray*}
For the $y$-RVP, we have
\begin{eqnarray*}
 \Sigma_0 &=& \frac{\Delta u_2^2}{2\pi^2v_1v_2|w|(w^2-1)}w(w-\lambda)l \ , \\
 \Sigma_1 &=& \frac{\Delta u_2^2}{2\pi^2v_1v_2|w|(w^2-1)}(w-\lambda)l \ ,
\end{eqnarray*}
and $\delta S_{\Gamma}=V_2\sigma_2$, where
\begin{eqnarray*}
 V_2 &=& -\frac{\Delta u_2^3}{4\pi^2v_1v_2|w|}l \ .
\end{eqnarray*}
Finally, for the RM,
\begin{eqnarray*}
 \Sigma_0 &=& \frac{\Delta u_3^2}{2\pi^2v_1v_2|w|(w^2-1)}w(w-\lambda)l \ , \\
 \Sigma_1 &=& \frac{\Delta u_3^2}{2\pi^2v_1v_2|w|(w^2-1)}(w-\lambda)l \ ,
\end{eqnarray*}
and $\delta S_{\Gamma}=0$. With the similar procedure, we obtain the one-loop RG equations in the main
text.



\begin{thebibliography}{99}
\bibitem{foot1} By setting $w=0$, we find that $d\gamma_{0l}/dl=0$ when $\gamma_{0l}=0$ and
        $d\gamma_{1l}/dl=0$ when $\gamma_{1l}=0$. This implies that we can treat the RSP and the $x$-RVP
        separately for untilted DSMs. Moreover, for untilted DSMs, the RSP is a relevant perturbation,
        while the $x$-RVP is a marginal perturbation. This is identical to the previous results in Ref.
        \onlinecite{Ludwig}. For the untilted DSM, $d\lambda_l/dl\propto\lambda_l$ when either
        $\gamma_{0l}=0$ or $\gamma_{1l}=0$. Thus, it is not necessary to include the term
        $\psi^{\dagger}_{\xi\sigma}\sigma_1\partial_{\tau}\psi_{\xi\sigma}$.
\bibitem{foot2} In fact, there is another solution for Eq. (\ref{wf2rgsol1}): $\lambda_l\rightarrow+\infty$.
        However, this solution cannot be reached for the initial value $\lambda=0$.
\end{thebibliography}
\end{document}